\documentclass[Journal,letterpaper,InsideFigs,NoLineNumbers]{ascelike-new}
\usepackage[utf8]{inputenc}
\usepackage[T1]{fontenc}
\usepackage{lmodern}
\usepackage{graphicx}
\usepackage{longtable}
\usepackage[figurename=Fig.,labelfont=bf,labelsep=period]{caption}
\usepackage{subcaption}
\usepackage{amsmath}
\usepackage{newtxtext,newtxmath}
\usepackage[colorlinks=true,citecolor=red,linkcolor=black]{hyperref}
\NameTag{Sac-Morane, \today}
\begin{document}

% You will need to make the title all-caps
\title{IMPORTANCE OF PRECIPITATION ON THE SLOWDOWN OF CREEP BEHAVIOR INDUCED BY PRESSURE-SOLUTION}

\author[1]{Alexandre Sac-Morane}
\author[2]{Hadrien Rattez}
\author[3]{Manolis Veveakis}

\affil[1]{PhD. Candidate. Multiphysics Geomechanics Lab, Duke University, Hudson Hall Annex, Room No. 053A, Durham 27708, NC, USA. alexandre.sacmorane@duke.edu}
\affil[2]{Assistant Professor. Institute of Mechanics, Materials and Civil Engineering, UCLouvain, Place du Levant 1, Louvain-la-Neuve 1348, Belgium. hadrien.rattez@uclouvain.be}
\affil[3]{Associate Professor. Multiphysics Geomechanics Lab, Duke University, Hudson Hall Annex, Room No. 053A, Durham 27708, NC, USA. manolis.veveakis@duke.edu}

\maketitle

%%=======================================================%%

% Please include an abstract:
\begin{abstract}
Pressure-solution is a chemo-mechanical process, involving dissolution at grain/asperity contacts and precipitation away from them. It induces a compaction in time of rocks and sediments. The present study investigates numerically the impact of precipitation on the slowdown of creep behavior induced by pressure-solution. A recently published framework, called the Phase-Field Discrete Element Model, is carefully calibrated against existing indentation experiments and validated for other rate-limiting scenarios. It is shown that when precipitation is relatively slow, the slowdown of pressure-solution is due to a chemical mechanism (accumulation of solute concentration within the pore space), whereas, at fast precipitation, the slowdown is due to a mechanical mechanism (stress reduction at the contact).
\end{abstract}

%%=======================================================%%

\section{Introduction}

\begin{figure}[h]
    \centering
    \includegraphics[width=0.6\linewidth]{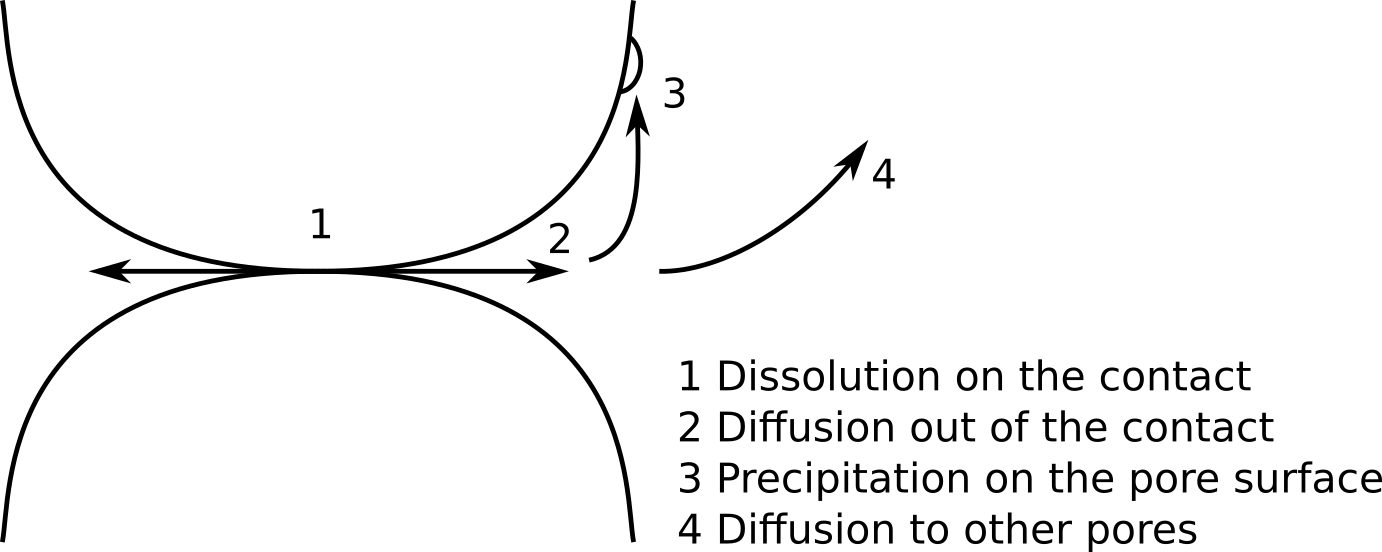}
    \caption{Scheme of the pressure-solution phenomenon.}
    \label{Pressure Solution Scheme}
\end{figure}

Pressure-solution creep is a common phenomenon observed in porous geomaterials undergoing diagenesis or involved in earthquake nucleation \cite{Weyl1959,Rutter1976,Angevine1982,Tada1989,Yasuhara2005}.
The phenomenon includes three fundamental chemo-mechanical processes at the microscale: (i) dissolution due to stress concentration at grain contacts, (ii) diffusive transport of dissolved mass from the contact to the pore space, and (iii) precipitation of the solute on the less stressed surface of the grains. Fig. \ref{Pressure Solution Scheme} provides a schematic representation of the pressure-solution process at the boundaries between two grains.
Competition exists between the three fundamental processes (dissolution, diffusion, precipitation) for the pressure-solution rate. If one is much slower than the others, the associated kinetic is the rate-limiting process \cite{Rutter1976,Raj1982}, imposing its pace on the global phenomenon. On the other hand, if the kinetics of the three processes are similar, more complex relationships need to be considered. Furthermore, the rate-limiting process may change as the configuration evolves during the pressure-solution phenomenon. 
The principal consequences for the porous material are a time-dependent compaction (creep) and a pore structure evolution \cite{Renard2004,Zubtsov2004,Schwichtenberg2022}. This reduction of the porosity is attributable to a mechanical process (grain reorganization) and a chemical one (pore filling by the diffusive mass transfer). 

In the literature, pressure-solution is described using mathematical models at the continuum scale based on experimental data \cite{Rutter1976,Raj1982,Spiers1990,Gratier2009,Zhu2015} or using micro-mechanical models with simplifying assumptions \cite{Spiers1990b,Guevel2020,Lu2021}. 
However, a comprehensive understanding of this complex phenomenon for granular materials necessitates the capture at the microscale of at least two key aspects: the granular reorganization and the evolution of particles' shapes. Indeed, these two aspects exert a significant influence on the force transmitted in the granular media, and as stress concentration at the contacts is the driving force for pressure-solution, it should strongly influence pressure-solution rate. Granular reorganization is not considered in the existing models at the microscale \cite{Spiers1990b,Guevel2020,Lu2021} and the precipitation (influencing the grain shape evolution) is not modeled in \cite{Spiers1990b,Lu2021}. 
Recently, a coupling between a Phase-Field (PF) model and a Discrete Element Model (DEM) \cite{SacMorane2024} has been developed, allowing to capture: (i) granular reorganization, (ii) irregular grain shape, (iii) force transmitted at the contacts, (iv) heterogeneous dissolution/precipitation, (v) diffusion of the solute in the pore space and (vi) rate-limiting processes. 

The objective of this paper is to present improvements to the PFDEM framework described in \cite{SacMorane2024} and to show the capabilities of the model to describe pressure-solution at the microscale. In particular, the contact law between particles used in the DEM has been modified, from an overlap-based approach to a volume-based one. This point is fundamental to face undesirable effects, such as the fact that a contact force can be the same for two contacts with a similar overlap but different overlapping volumes \cite{vanderHaven2023,Feng2023}. Furthermore, the previous distinct mechanical and chemical destabilization terms in \cite{SacMorane2024} have been combined into one coupled term, more common in the literature on pressure-solution \cite{Paterson1973,Lu2021}.
These modifications are derived in the following Sections.

This new formulation is calibrated using previous indentation experiments conducted on quartz \cite{Gratier2009}. Even if most of the parameters are well-known in the literature, two of them remain poorly constrained: the diffusivity and the dissolution/precipitation kinetics coefficient. Indeed, the diffusion coefficient obtained by interpolation in \cite{Gratier2009}, varies by several orders of magnitude with the model used to fit the noisy data. Subsequently, a validation is carried out, showing that the two well-known rate-limiting scenarios (diffusion or dissolution rate-limiting) \cite{Rutter1976,Raj1982} are well reproduced. Indeed, the different parameter dependencies for the kinetic of the indenter test must be verified even if the calibration campaign has been conducted.
Finally, the influence of precipitation on the slowdown of the creep behavior is explored through a granular configuration.

%%=======================================================%%

\section{Calibration and validation of the model with indentation experiments}
\label{Indentation Section}

The purpose of this Section is to calibrate the phase-field parameters of the model described in \cite{SacMorane2024} and then to validate the formulation. First, the phase-field model  \cite{Landau1936,Cahn1958,Allen1979} is presented. In the present study, the tilting energy $e_d$, see Equation \ref{Tilting Term}, has been reformulated compared to \cite{SacMorane2024}. The chemical and mechanical terms are separated in \cite{SacMorane2024}, a coupled one is used here, more common in the literature on pressure-solution \cite{Paterson1973,Lu2021}.
The formulation is applied to model indentation experiments described in \cite{Gratier2009,Gratier1993,Dysthe2002,Gratier2014} and the calibration is done by reproducing the data from \cite{Gratier2009}. In fact, some parameters (diffusivity and dissolution kinetic) are still not well constrained.
The formulation is then verified by comparison with the two well-known macro-models for the compaction rate: the diffusion and the dissolution rate-limiting \cite{Rutter1976,Raj1982}. 

%%===========================%%

\subsection{Phase-field formulation}
\label{Phase Field Section}

A granular material consists of grains and pores. This porous material is represented here in the phase-field space by a combination of phase variables (one phase per grain). This phase variable equals $1$ if it is inside the corresponding grain and equals $0$ otherwise. 
The phase-field method, specifically the Allen-Cahn equation described in Equation \ref{AC formulation}, is employed to describe the dissolution/precipitation of the material \cite{Allen1979}. In this approach, non-conserved order parameters $\eta_j$ are used to represent the phase transformation. 

\begin{equation}
\frac{\partial \eta_j}{\partial t} = -L_j\left(\frac{\partial \left(f_{loc}+E_d\right)}{\partial \eta_j} - \kappa_j \nabla^2 \eta_j \right)
\label{AC formulation}
\end{equation}

\begin{figure}[ht]
    \centering
    \includegraphics[width=0.7\linewidth]{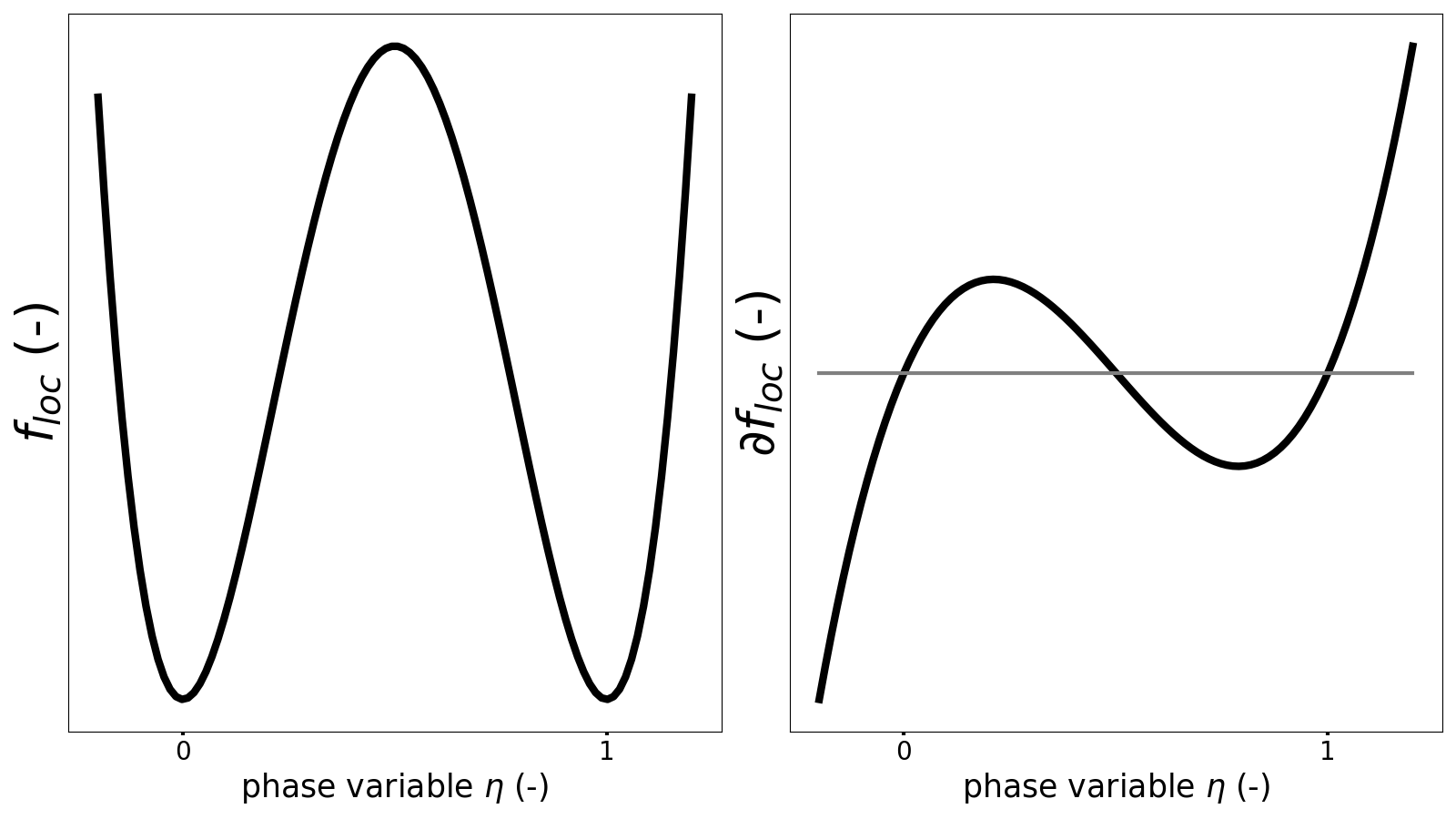}
    \caption{A double-well function of the phase variable $\eta$ as a local free energy density $f_{loc}(\eta)$ and its derivative $\frac{\partial f_{loc}}{\partial\eta}(\eta)$.}
    \label{Free Energy Figure}
\end{figure}

In this equation, the terms $L_j$ and $\kappa_j$ correspond to the order parameter mobility and the gradient energy coefficient, respectively. The order parameter mobility affects the overall dissolution/precipitation kinetic, while the gradient energy coefficient influences the interface width and the reaction kinetic.
The term $E_d$ encompasses additional energy sources introduced into the system, such as mechanical, chemical, or thermal loading. The local free energy density $f_{loc}$ is specified in Equation \ref{Free energy} and is represented in Fig. \ref{Free Energy Figure}.

\begin{equation}
f_{loc} = h \times \left(\sum_j \eta_j^2(1-\eta_j)^2\right)
\label{Free energy}
\end{equation}

The local free energy density $f_{loc}$ employs a double-well function with a barrier height $h$ \cite{Takaki2014}. Notice that the minima of this potential energy are located at $\eta_j=0$ and $\eta_j=1$. At the equilibrium and without external destabilization, the phase variables stay at $\eta_j=0$ or $\eta_j=1$ (no dissolution/precipitation).

\begin{figure}[ht]
    \centering
    a) \includegraphics[width=0.45\linewidth]{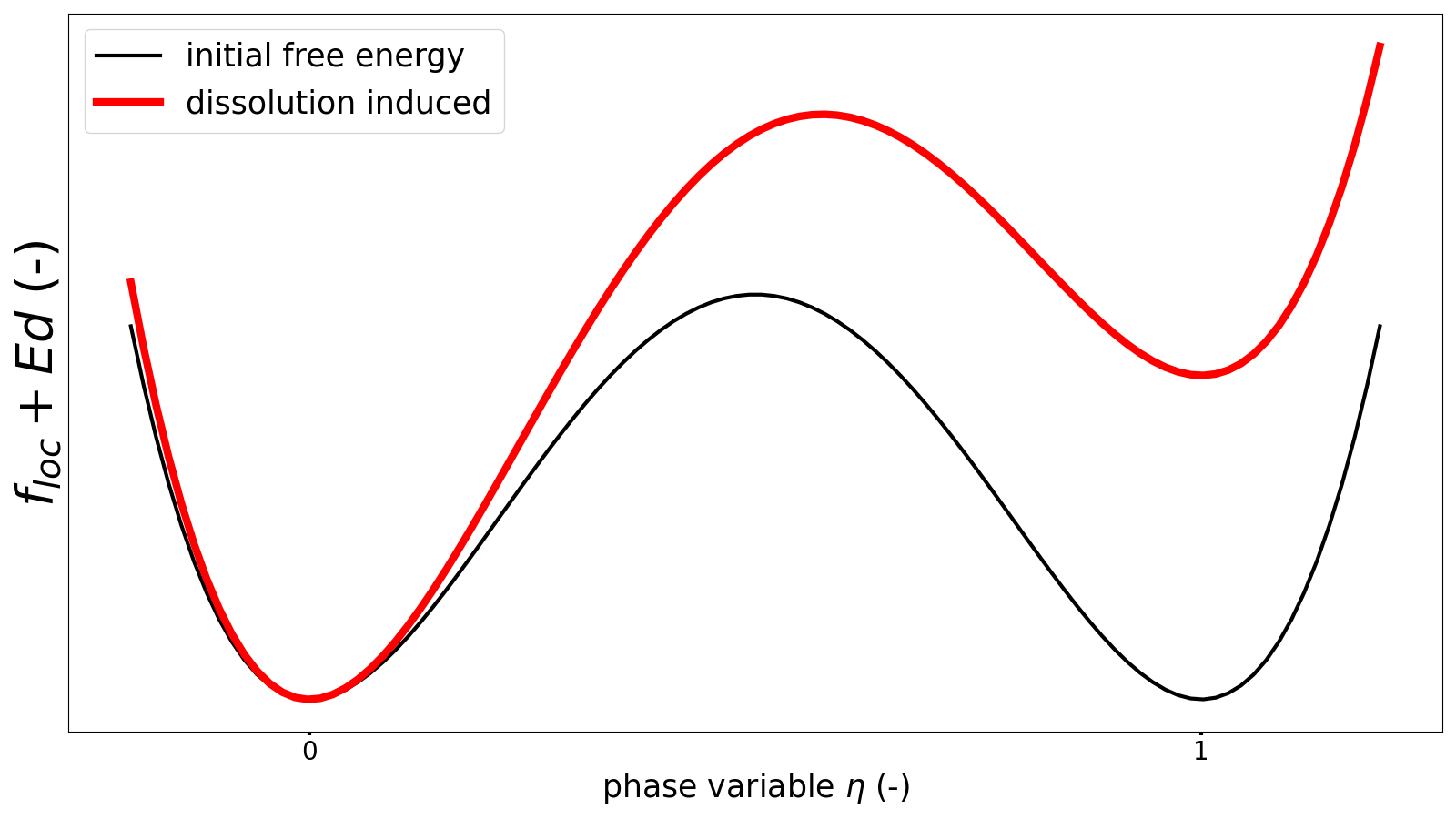}~
    b) \includegraphics[width=0.45\linewidth]{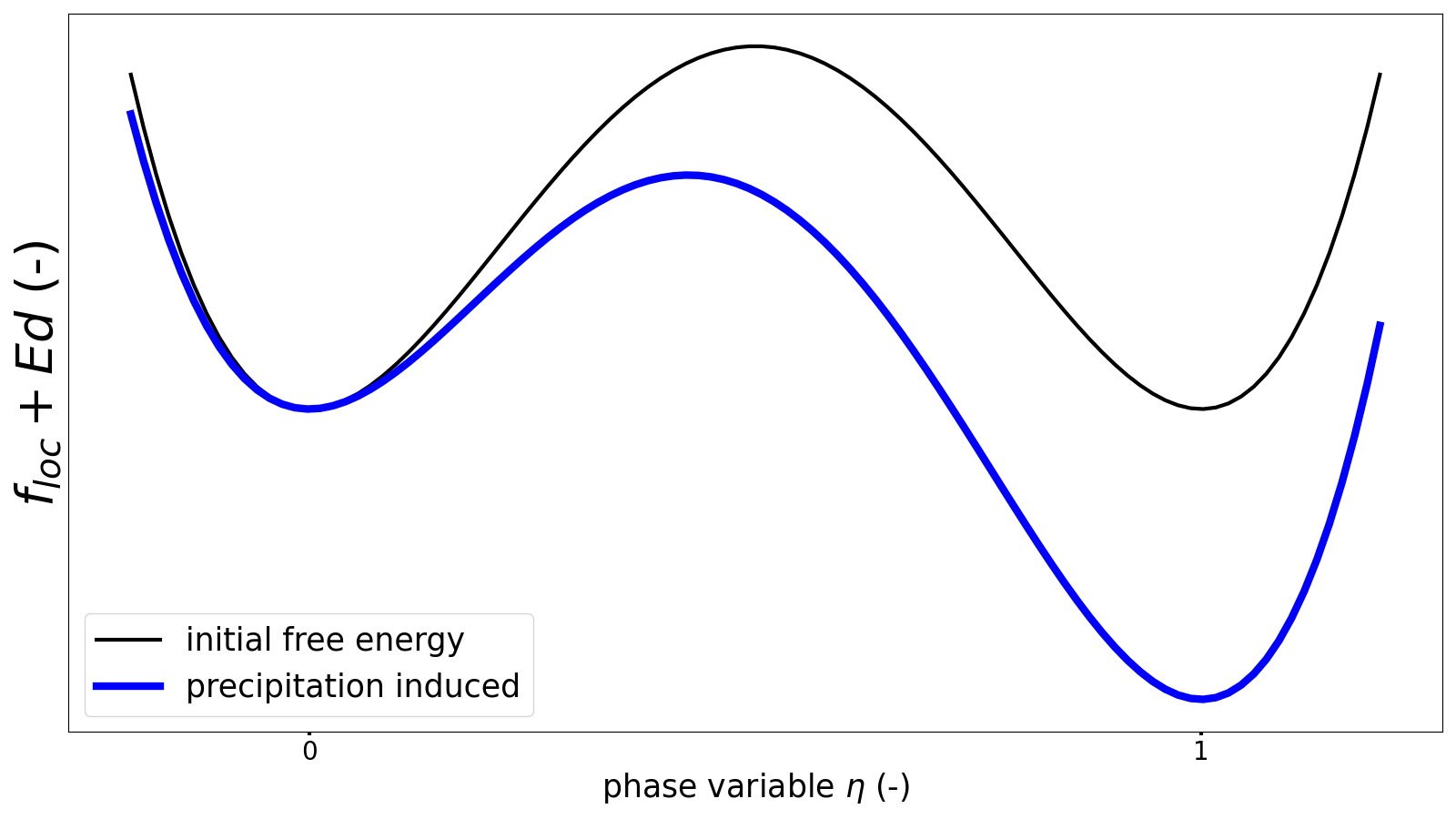}
    \caption{The double-well function is destabilized to favor a) dissolution or b) precipitation.}
    \label{Destabilized Double Well}
\end{figure}

To obtain localized dissolution/precipitation, an external source term $E_d$ is added to tilt the initial free energy to favor dissolution/precipitation, as depicted in Fig. \ref{Destabilized Double Well} and Equation \ref{Tilting Term}. 
When the dissolution phenomenon dominates, the double-well function is tilted towards $\eta_j = 0$, resulting in the dissolution of the material. Conversely, when the precipitation phenomenon dominates, the double-well function is tilted towards $\eta_j = 1$, indicating the material's tendency to precipitate.
The amplitude of the tilting affects the kinetic of the dissolution/precipitation.

Before formulating $E_d$, another variable, denoted as $c$, is introduced in the phase-field formulation to model solute concentration in the pore fluid. The Equation \ref{AC formulation} is coupled with the diffusion equation for the variable $c$, as given in Equation \ref{Solute formulation}.

\begin{equation}
	\frac{\partial c}{\partial t} = -\sum\limits_j\left(\beta_j\frac{\partial \eta_j}{\partial t}\right) + \kappa_c \nabla^2 c    
    \label{Solute formulation}
\end{equation}
The Equation \ref{Solute formulation} represents the conservation of solute mass and is solved using the variations of phase-field variables $\eta_j$ as a source/sink term. $\beta_j$ is a conversion factor between $\eta_j$ and $c$, and $\kappa_c$ represents the gradient coefficient that controls the diffusion rate. Notably, the gradient coefficient $\kappa_c$ is assumed to be 0 within the grains, and nonzero in the contact and the pore. A water film is assumed here at the contact between grains \cite{vanNoort2008,Gratier2009}. Furthermore, the gradient coefficient is larger in the pore (free fluid) than at the contact (trapped fluid) \cite{vanNoort2008}, an arbitrary factor $100$ could be applied.
The solute $c$ diffuses from the contact area, where dissolution is triggered, into the pore space. 
The algorithm's implementation for building the heterogeneous gradient energy coefficient $\kappa_c$ is described in Appendix B from \cite{SacMorane2024}.
The terms $\frac{\partial c}{\partial t} = - \sum\limits_j \beta_j \frac{\partial \eta_j}{\partial t}$ in Equation \ref{Solute formulation} ensure the conservation of mass. When a grain $\eta_j$ dissolves, solute concentration $c$ is generated, and vice versa, when matter $\eta_j$ is generated through precipitation, the solute concentration $c$ decreases.

The tilting term $E_d=e_d\times \sum\limits_j \left(3\eta_j^2-2\eta_j^3\right)$ is defined in Equation \ref{Tilting Term}. This term has been modified here compared to \cite{SacMorane2024}. Indeed, in the previous model two distinct terms were assumed, one for the chemistry and one for the mechanics. This coupled term $e_d$ models the fact that the chemical potential is directly affected by the stress transmitted in between grains as the two mechanisms are deeply connected. This formulation is closer to the mechanism described for pressure-solution in \cite{Weyl1959,Paterson1973,Lu2021}. 

\begin{equation}
e_d = k_{diss/prec} \times a_s \times \left(1-\frac{c}{c_{eq}\,a_s}\right)
\label{Tilting Term}
\end{equation}

The external energy depends especially on the solid activity $a_s$ and the solute concentration $c$, $c_{eq}$ representing the solute concentration at the equilibrium without any external destabilization. The solid activity $a_s$ is defined in Equation \ref{Solid Activity Equation} \cite{Paterson1973,Lu2021}. The coefficient $k_{diss/prec}$ is related to the kinetic of the dissolution/precipitation chemical reaction. The larger it is, the larger the tilting is, and the faster the dissolution/precipitation is. If $c<c_{eq}a_s$, dissolution occurs, $k_{diss/prec} = k_{diss}$. Conversely, if $c>c_{eq}a_s$, precipitation occurs, $k_{diss/prec} = k_{prec}$.

\begin{equation}
    a_s = \text{exp}\left(\frac{P\times V_m}{RT}\right)
    \label{Solid Activity Equation}
\end{equation}
where, $P$ is the pressure applied, $V_m$ is the molar volume, $R$ is the 
gas constant and $T$ is the temperature (assumed constant here). 

It seems that the solid activity $a_s$ increases in proportion to the stress transmitted at the contact $P$. This modification affects the value of the solute concentration at the equilibrium $c_{eq}\,a_s$. The established chemical equilibrium in the fluid film ($c \text{ versus } c_{eq}\,a_s$) becomes unverified. The material dissolves in the contact zone to reach the chemical equilibrium $c = c_{eq}\,a_s$. A gradient in the solute concentration values appears between the contact zone $c = c_{eq}\,a_s$ and the pore space $c = c_{eq}$. This gradient gives rise to the diffusion of the solute concentration. Subsequently, the established chemical equilibrium in the pore space ($c \text{ versus } c_{eq}$) becomes unverified. The material precipitates in the pore space to reach the chemical equilibrium $c=c_{eq}$.

Equations \ref{AC formulation} and \ref{Solute formulation} are solved on a microstructure geometry with the open-source finite element software MOOSE \cite{Moose2020}. This software uses automatic differentiation to solve the system of equations with an implicit formulation.  

%%===========================%%

\subsection{Numerical model}
\label{Calibration Numerical Model}

\begin{figure}[h]
    \centering
    \includegraphics[width=0.6\linewidth]{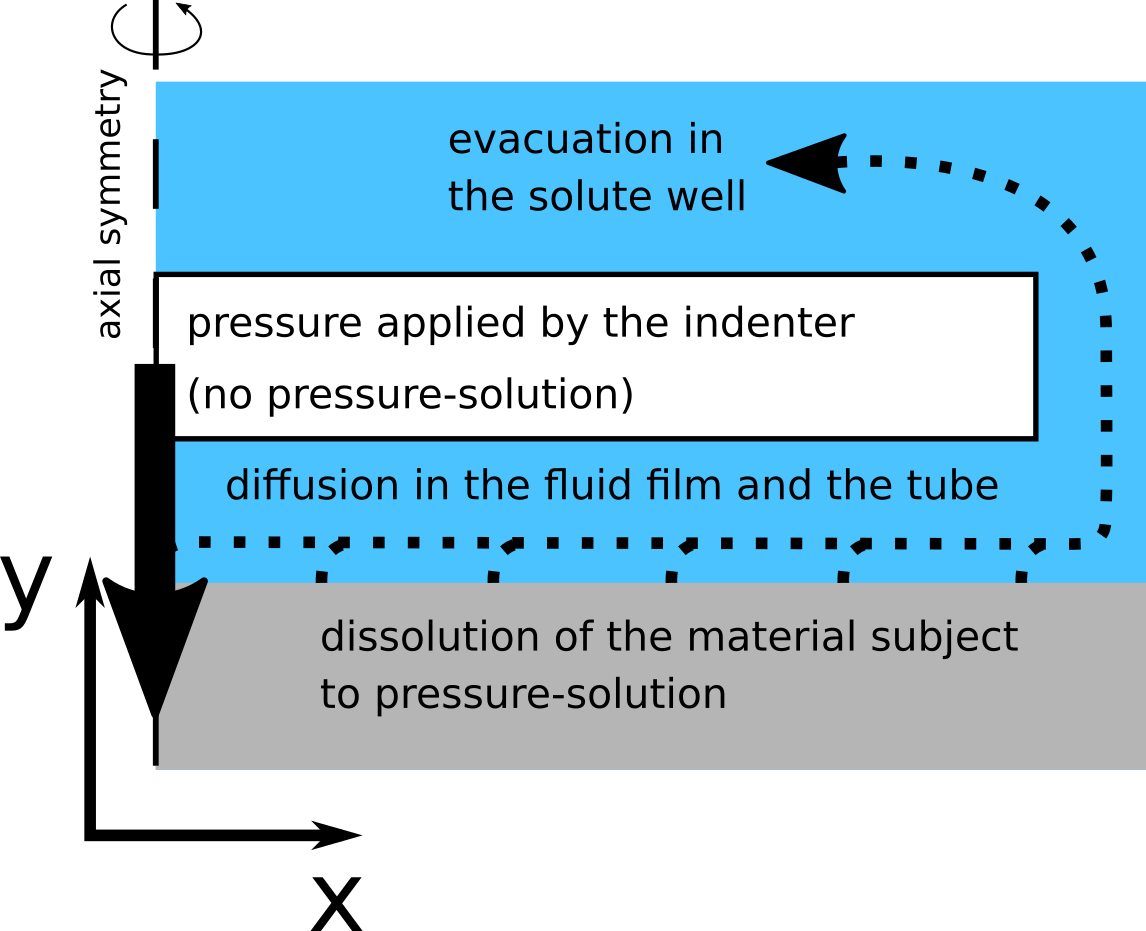}
    \caption{Scheme of the indentation experiment composed of an indenter and a material subject to the pressure-solution phenomenon. The material is dissolved at the contact surface, diffuses into the fluid film and is evacuated in the well through a tube.}
    \label{Scheme Indenter}
\end{figure}

The framework employed for parameter calibration and formulation validation is presented herein. A schematic representation of an indentation experiment \cite{Gratier1993,Dysthe2002,Gratier2009,Gratier2014} is presented in Fig. \ref{Scheme Indenter}. 
A material subjected to pressure-solution is situated beneath an indenter. This indenter exerts a constant pressure. It should be noted that this indenter is not subject to pressure-solution. 
The material dissolves at the contact. Subsequently, it diffuses in the fluid film,  the tube and the well. Finally, the solute is evacuated from the well.

Despite the existence of disparate discourses pertaining to the structural configuration of this film \cite{vanNoort2008}, it is postulated that a thin fluid film exists between the indenter and the material \cite{Gratier2009}. Indeed, it could be a film, an island-channel network or a film with cracks. Experimental evidence indicates that the film in question is approximately 2–10 nm in thickness, corresponding to a water molecule layer of 8–40 molecules \cite{Gratier2009}. From a numerical standpoint, it is not feasible to respect this size, particularly compared to the size of the indenter. Subsequently, the film width is artificially increased. Nevertheless, the coefficient of diffusion is decreased in proportion. The minimum size of this film is related to the phase-field interface width. Given that material dissolution occurs within this zone, it is necessary for the film to contain it.

Once the solute has reached the reservoir at the top, it is evacuated at a specified frequency. The frequency in question is of great consequence with regard to the quality of the results. Indeed, an increase in solute concentration within the well may result in a slowing of diffusion as the gradient diminishes, see Appendix I. The dimension of the well and the frequency of updates to the simulation can be modified in order to regulate the solute concentration within the well. In the following results, an effort has been made to compare simulations with the same solute concentration in the well.

The phase-field parameters are linked, especially through Equation \ref{PF Parameters link} \cite{Takaki2014}.

\begin{equation}
    \kappa = h\frac{144\Delta x^2}{b^2}
    \label{PF Parameters link}
\end{equation}
where, $\kappa$ is the gradient energy coefficient (from Equation \ref{AC formulation}), $h$ is the barrier height (from Equation \ref{Free energy}), $\Delta x$ is the mesh size (a phase-field interface of the size of $6 \Delta x$ is assumed), and $b=2.2$ \cite{Takaki2014}.
A strong mesh dependency exists because of the phase-field definition, see Equation \ref{PF Parameters link}.  Appendix I presents a method to cancel this influence.

It is important to notice that a constant pressure is applied along the indenter surface. However, the normal stress appears larger at the center of the contact than at the side \cite{Weyl1959,Rutter1976,Gratier2009}. A control factor $f_{control}$ is applied to force the material interface to stay as flat as possible, see Equation \ref{Control Front}. 

\begin{equation}
    a_{s,i} = a_s \times \left(y_{front,i}-\overline{y_{front}}\right)\times f_{control}
    \label{Control Front}
\end{equation}
where, $a_{s,i}$ is the corrected solid activity applied in Equation \ref{Tilting Term}, $a_{s}$ is the solid activity defined in Equation \ref{Solid Activity Equation}, $y_{front,i}$ is the coordinate of the front and $\overline{y_{front}}$ is the mean coordinate of the front along the indenter.

%%===========================%%

\subsection{Calibration of the model}
\label{Calibration Section}

An initial set of simulations is conducted to calibrate the model with the data from the experiments presented in \cite{Gratier2009}. A movie example is provided as supplemental material. The majority of the parameters are known, including the equilibrium concentration $c_{eq}$, the molar volume $V_s=1/\beta_j$, the gas constant $R$, the temperature $T$ or the indenter diameter $d$ \cite{Gratier2009}. They are presented in Table \ref{Parameter Known}. The sole unknown parameter is the coefficient $\kappa_c\times w$, where $\kappa_c$ is the diffusion coefficient of the solute in the film and $w$ is the film width. It should be noted that the estimation provided in \cite{Gratier2009} appears to be dependent on the fitting method employed, with significant discrepancies observed between the reported values. For instance, the variation in the estimated coefficient, when fitted using either the proportionality ($\dot{y} \propto P$) or the power law ($\dot{y} \propto P^n$) models, is of a magnitude that spans several orders of magnitude. A trial-and-error method was employed to obtain simulated kinetics in the experimental data cloud. 
Moreover, as the sample is diffusion rate limited \cite{Gratier2009}, it is not possible to determine the dissolution kinetic coefficient $k_{diss}$ from the experiments. A numerical criterion is employed to estimate this parameter, $\overline{sat_{film}}\geq80\%$, where $\overline{sat_{film}}$ is the mean saturation in the fluid film defined as $sat_{film}=\frac{c-c_{eq}}{a_s c_{eq}-c_{eq}}$. 
The remaining parameters from Table \ref{Parameter Known} are numerical parameters, such as the barrier energy $h$, the mobility $L_j$, the mesh size $\Delta x$, the film width $w$ and the well saturation $sat_{well}$. They are employed for the phase-field description and to define the update frequency, see Appendix \ref{Preliminary Frequency Well Size Dependency}.

\begin{table}[h]
    \centering
    \begin{tabular}{|l|r|}
        \hline
        Equilibrium concentration $c_{eq}$ & $0.73\times10^3\; \text{mol.m}^{-3}$  \\
        Molar volume $V_s=1/\beta_j$ & $2.2\times10^{-5}\;\text{m}^3\text{.mol}^{-1}$\\ 
        Gas constant $R$ & $8.32\;\text{J.mol}^{-1}\text{.K}^{-1}$\\
        Temperature $T$ & $623\;\text{K}$\\
        Indenter diameter $d$ & $200\times 10^{-6}\;\text{m}$\\
        \hline
        Barrier energy $h$ & $1$\\
        Mobility $L_j$ & $1\;\text{s}^{-1}$\\ 
        Mesh size $\Delta x$ & $d/400$\\
        Film width $w$ & $15\,\Delta x$\\
        Well saturation $sat_{well}=\frac{c_{well}-c_{eq}}{a_s c_{eq}-c_{eq}}$& $30\;\text{\%}$\\
        \hline
        Diffusivity $\kappa_c\; \times $ film width $w$& $4e^{-14}\;\text{m}^3\text{.s}^{-1}$\\
        Dissolution kinetic $k_{diss}=10\,k_{prec}$ & $0.015/\Delta x$\\
        \hline
    \end{tabular}
    \caption{Parameters used. The solute diffusion $\kappa_c$ is obtained by calibration and the dissolution kinetic $k_{diss}$ is adapted to verify the diffusion rate-limiting scenario.}
    \label{Parameter Known}
\end{table}

\begin{figure}[h]
    \centering
    \includegraphics[width=0.6\linewidth]{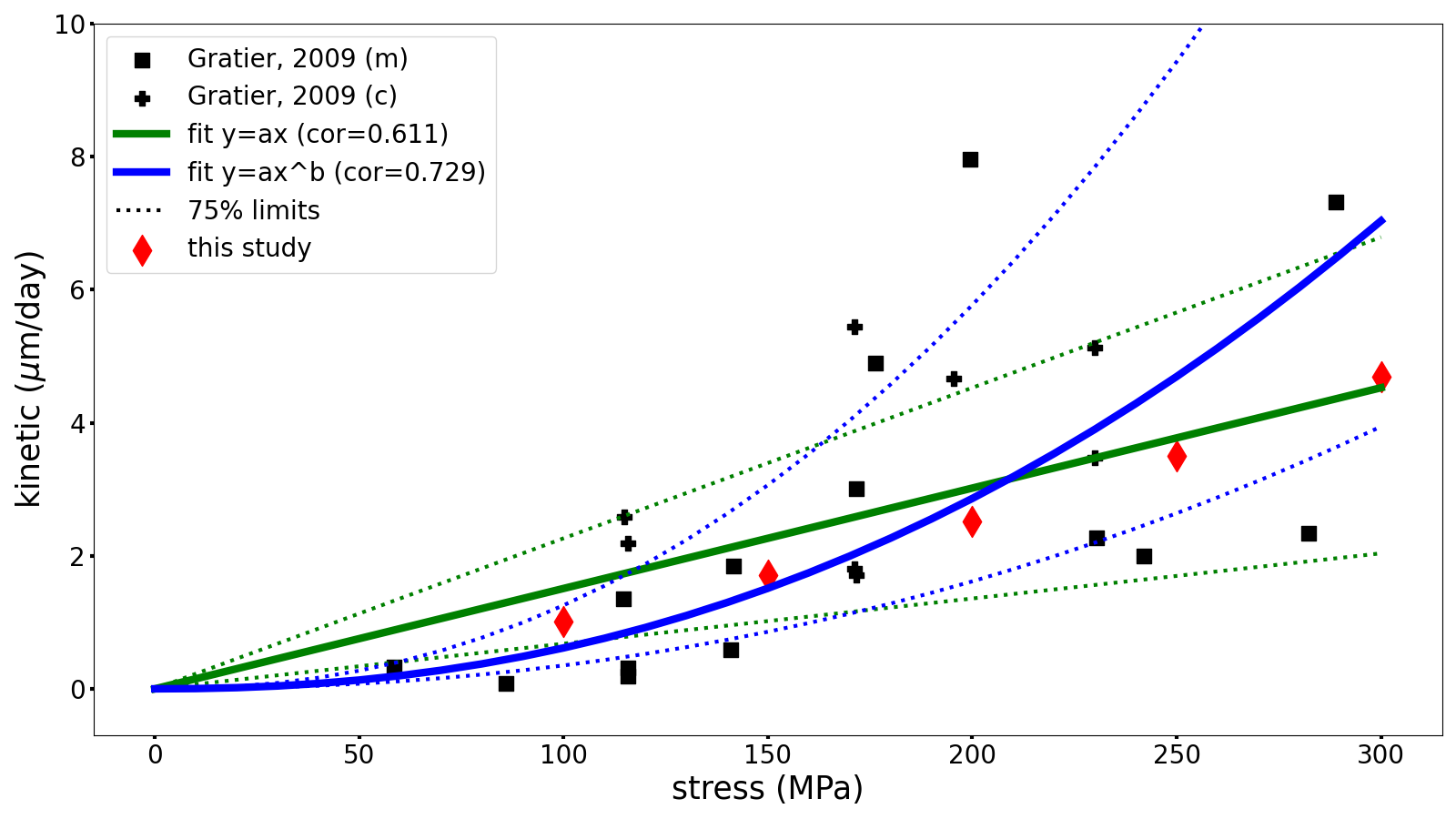}
    \caption{Comparison of the experimental and simulated kinetics of the indenter test at different confining pressures.}
    \label{Calibration Figure}
\end{figure}

The superposition of the results obtained with the data from \cite{Gratier2009} is presented in Fig. \ref{Calibration Figure}. Two fits have been performed on the data from \cite{Gratier2009} for comparison with the results obtained: $\dot{y}=a\times P$ and $\dot{y}=a\times P^b$. Subsequently, the various parameters of the fits were subjected to deterioration in order to ascertain the boundary limits. For example, the upper envelope for the $75\%$ limit is defined as $75\%$ of the data points from \cite{Gratier2009} being below it. Similarly, the lower envelope for the $75\%$ limit is defined as $75\%$ of the data points from \cite{Gratier2009} are larger. The results obtained with the phase-field description are in good agreement with the available data.

%%===========================%%

\subsection{Validation of the diffusion rate-limiting scenario}
\label{Validation Diffusion Section}

Despite the implementation of a calibration campaign, it is imperative to ascertain the accuracy of the model by replicating the different creep rate behaviors for pressure-solution associated with different parameters. Two well-known creep rate models exist, depending on the rate-limiting scenario \cite{Rutter1976,Raj1982}.

The diffusion rate-limiting one is defined as the diffusion kinetic process is the slowest \cite{Rutter1976,Raj1982,Spiers1990b,Gratier2009}. Numerically this scenario is verified by the criterion: $\overline{sat_{film}}\geq80\%$, where $\overline{sat_{film}}$ is the mean saturation in the fluid film. It is defined as $sat_{film}=\frac{c-c_{eq}}{a_s c_{eq}-c_{eq}}$, where $c$ is the solution concentration, $c_{eq}$ is the solution concentration at the equilibrium and $a_s$ is the solid activity defined Equation \ref{Solid Activity Equation}.
The displacement rate $\dot{y}$ during diffusion rate-limiting steady state is defined at Equation \ref{strain rate diffusion} \cite{Rutter1976,Raj1982,Spiers1990b,Gratier2009}.

\begin{equation}
    \dot{y}\propto \frac{\kappa_cw\, c_{eq}\, P^n}{d^2}
    \label{strain rate diffusion}
\end{equation}
where, $\kappa_c$ is the diffusion coefficient, $w$ is the film width, $c_{eq}$ is the solution concentration at the equilibrium, $P$ is the pressure applied, $n$ is an exponent applied to the pressure ($\approx 1-2$ \cite{Berest2019}) and $d$ is the diameter of the indenter ($\sim$ the distance to the free fluid). 

\begin{figure}[h]
    \centering
    a)\includegraphics[width=0.45\linewidth]{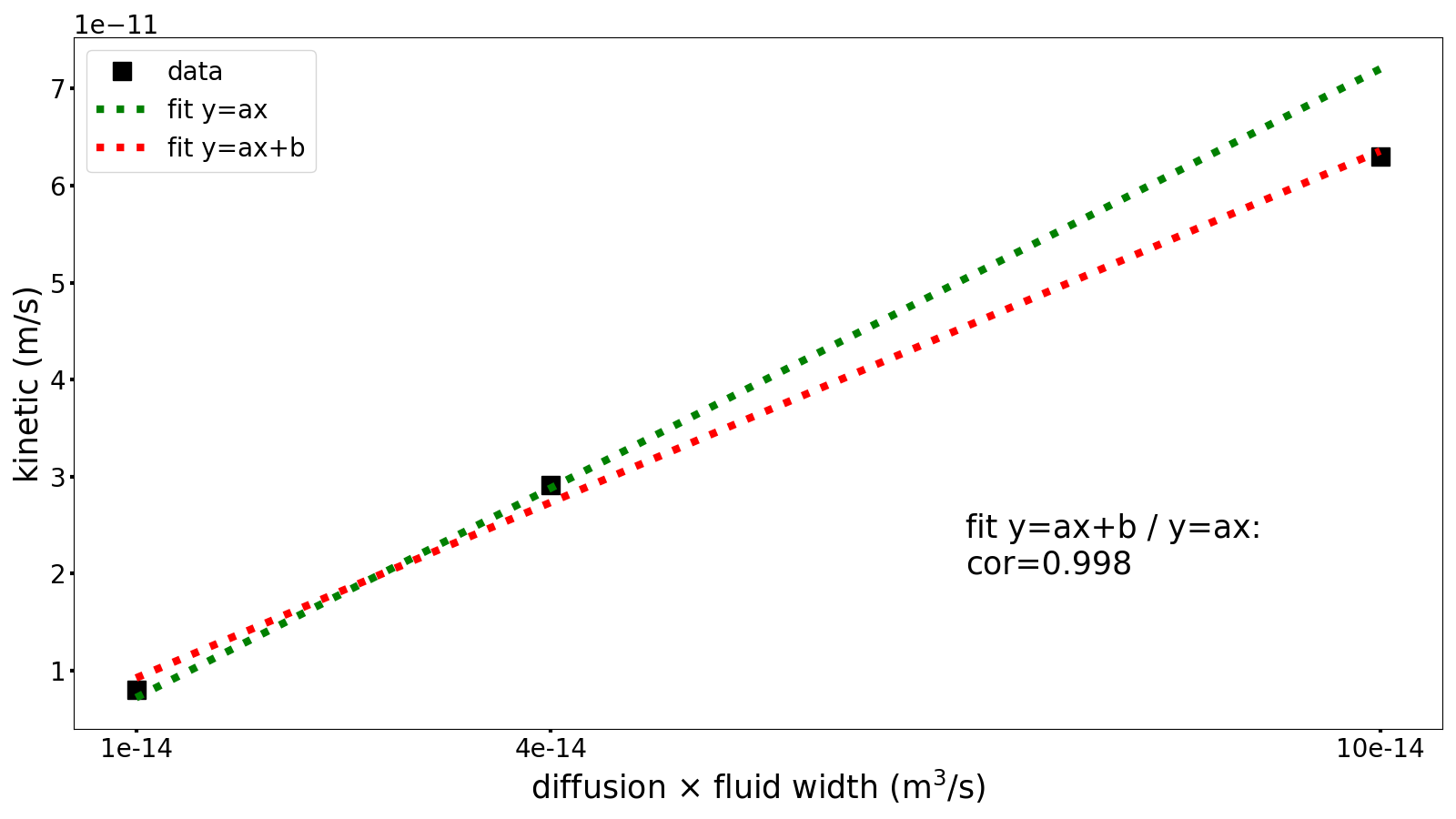}~
    b)\includegraphics[width=0.45\linewidth]{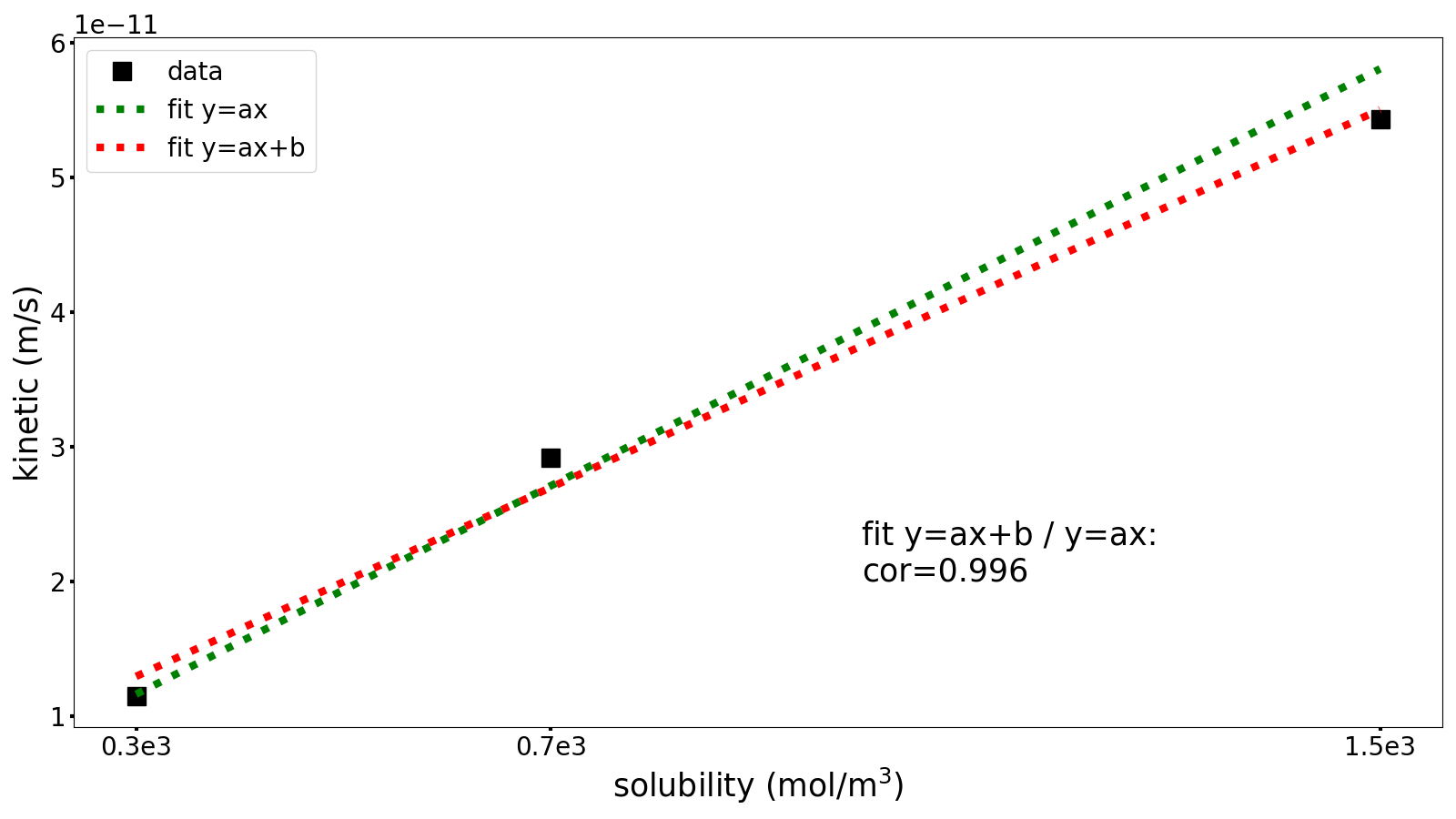}\\
    c)\includegraphics[width=0.45\linewidth]{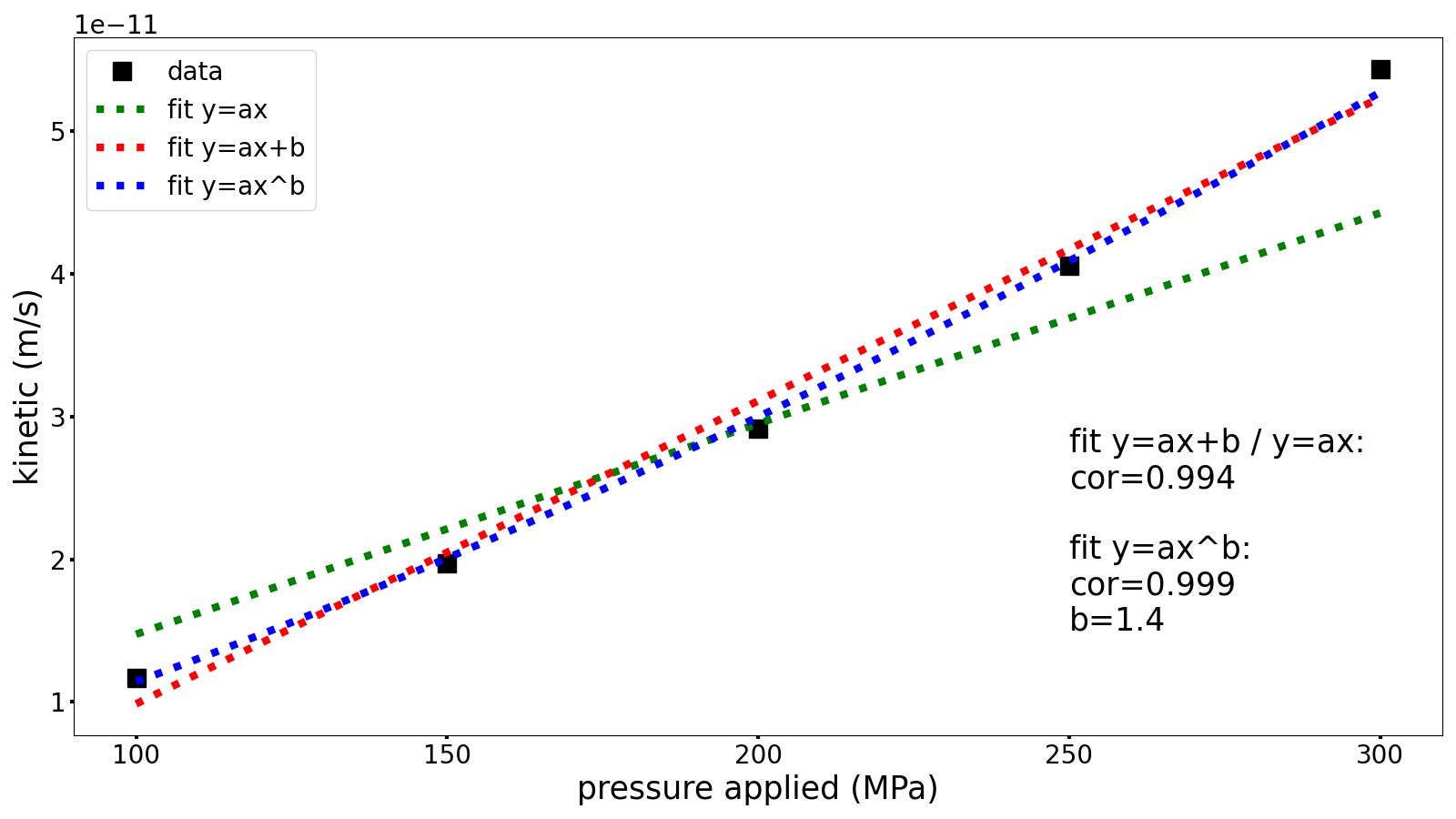}~
    d)\includegraphics[width=0.45\linewidth]{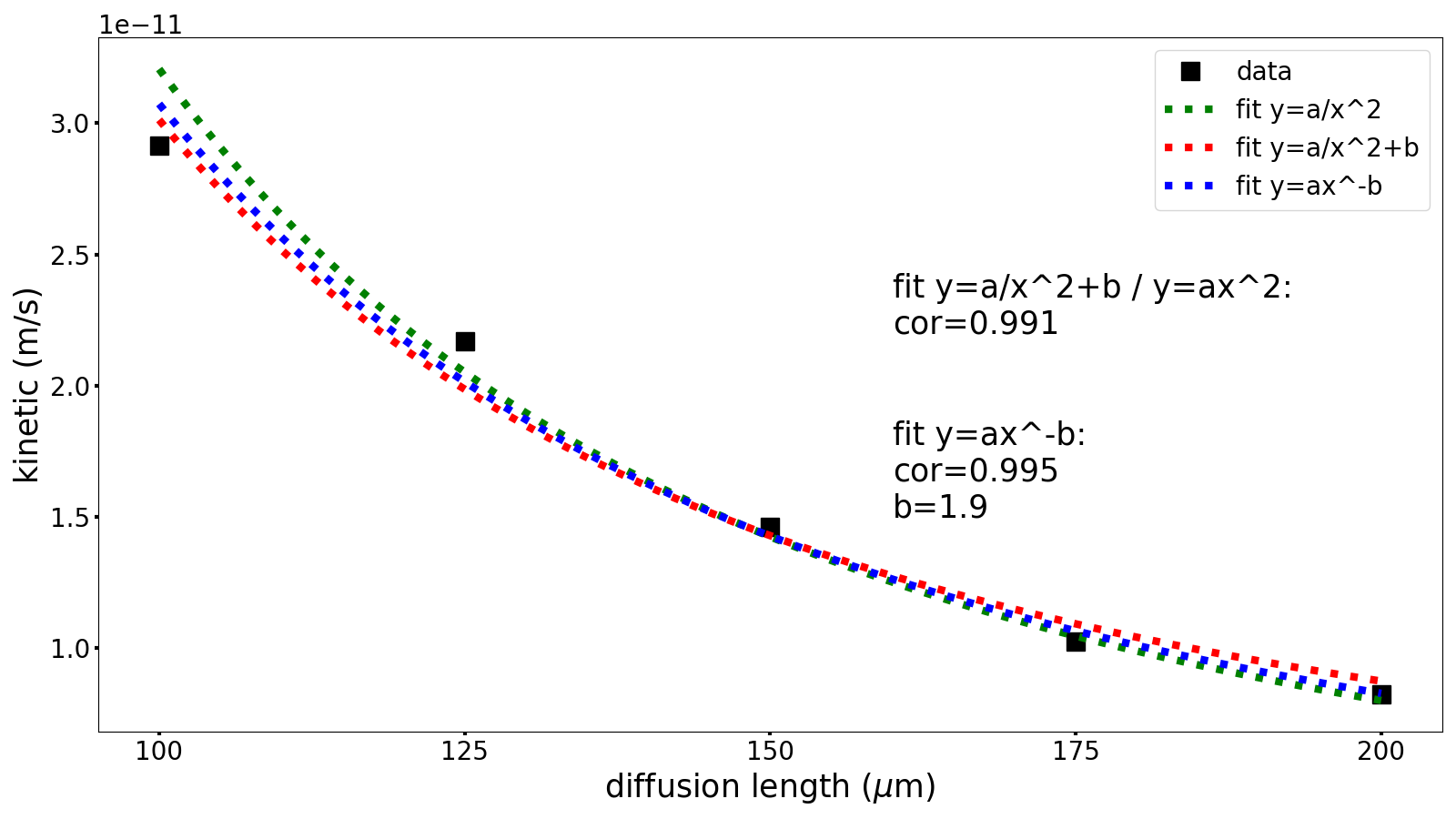}
    \caption{Comparison with the analytical model of the simulated influence on the kinetic of the indenter test of a) the diffusivity, b) the solubility, c) the pressure applied and d) the distance to the free fluid.}
    \label{Diff Rel Kin}
\end{figure}

\begin{figure}[h]
    \centering
    \includegraphics[width=0.6\linewidth]{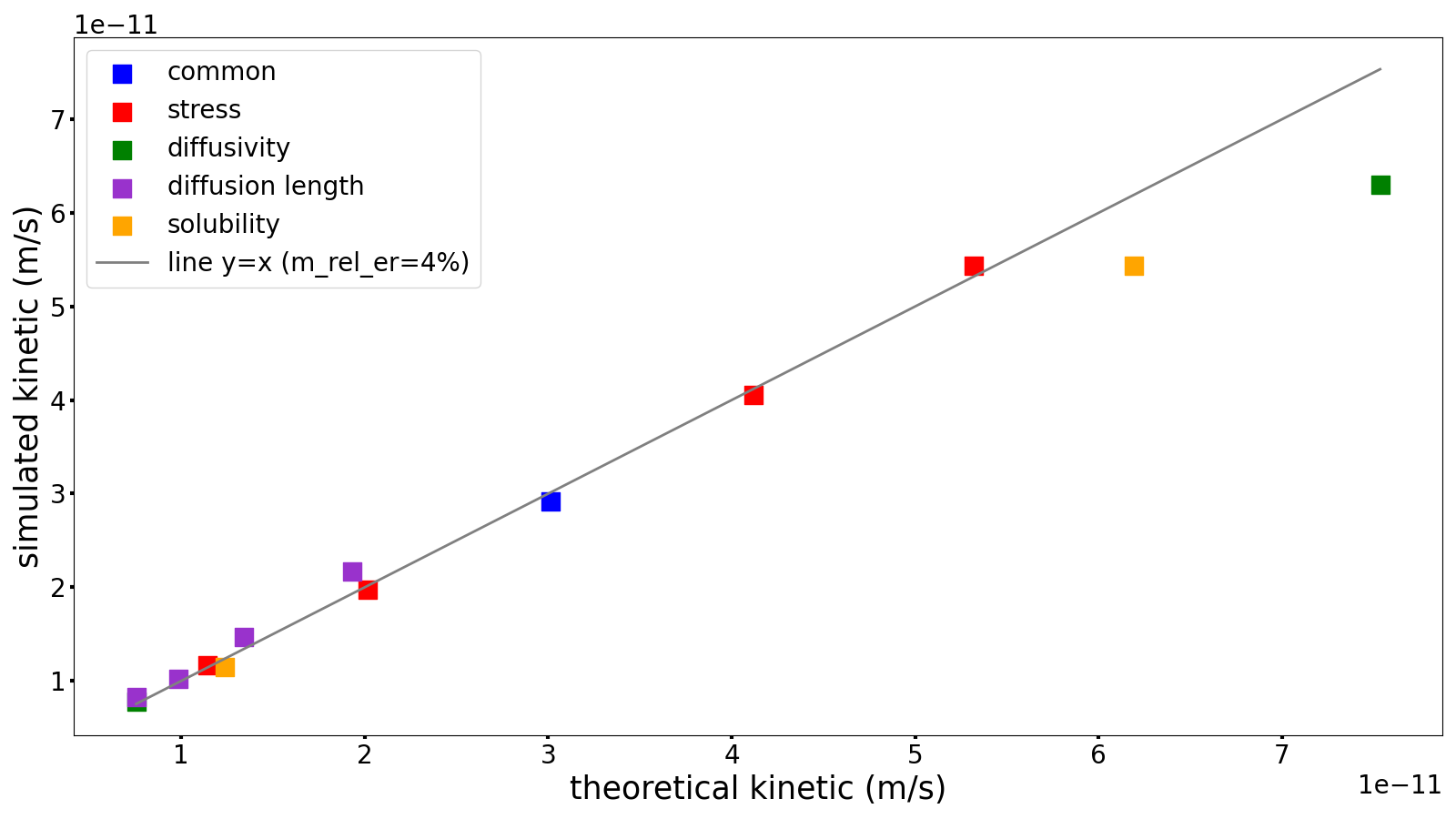}
    \caption{Comparison of the simulated and theoretical kinetics of the indenter test in the case of the diffusion rate-limiting.}
    \label{Diff de_sim de_the}
\end{figure}

A parametric investigation is done to verify the relation at Equation \ref{strain rate diffusion}. As shown in Table \ref{Parameter Indentation Diffusion}, the effect of each parameter is isolated: bold values represent the default ones (from \cite{Gratier2009}), whereas the others represent the ones used only in the specific investigation (domain definition from \cite{Gratier2009}).  Notice that the dissolution kinetic, related to $k_{diss}$ from Equation \ref{Tilting Term}, could have been modified to conserve the diffusion rate-limiting assumption $\left(\overline{sat_{film}}\geq 80\%\right)$. Results are given in Fig. \ref{Diff Rel Kin} and \ref{Diff de_sim de_the}.

\begin{table}[h]
    \centering
    \begin{tabular}{|l|r|}
        \hline
        diffusivity $\kappa_c$ $\times$ film width $w$& 1-\textbf{4}-10$e^{-14}$ m$^3$/s\\
        concentration at the equilibrium $c_{eq}$& 0.3-\textbf{0.73}-1.5$e^3$ mol/m$^3$\\
        pressure applied $P$& 100-150-\textbf{200}-250-300 MPa\\
        distance to the well $d$ & \textbf{100}-125-150-175-200 $\mu$m\\  
        \hline
    \end{tabular}
    \caption{Parameters used to verify Equation \ref{strain rate diffusion}. Bold values represent the default ones, whereas the others represent the ones used only in the specific investigation.}
    \label{Parameter Indentation Diffusion}
\end{table}

The influence of the diffusivity $\kappa_c$ on the strain rate $\dot{y}$ is explored in Fig. \ref{Diff Rel Kin}a. According to the Equation \ref{strain rate diffusion}, a proportional relation is expected. 
In agreement with it, data from Fig. \ref{Diff Rel Kin}a are well-fitted by a relation $\dot{y}=a\times \kappa_c$. Notice that the influence of the film width $w$ could also be explored. Remember that the film width $w$ is artificially increased compared to the reality to allow numerical simulations. The results would be similar to Fig. \ref{Diff Rel Kin}a if only the film fluid $w$ varies.

The influence of the solubility (through the solute concentration at the equilibrium $c_{eq}$) on the strain rate $\dot{y}$ is explored in Fig. \ref{Diff Rel Kin}b. It is anticipated that a proportional relationship will be identified, see Equation \ref{strain rate diffusion}. This factor has been observed experimentally \cite{Rutter1976,Zubtsov2005}.
In agreement with those observations, data from Fig. \ref{Diff Rel Kin}b are well-fitted by a relation $\dot{y}=a\times c_{eq}$. 
As explained sooner, an effort has been made to compare simulations with the same well saturation $sat_{well}$. The changing value $c_{eq}$ affects this parameter, as $sat_{well}=\frac{c-c_{eq}}{a_sc_{eq}-c_{eq}}$. A well-fitted linear relation is obtained.

The influence of the pressure applied $P$ on the strain rate $\dot{y}$ is explored in Fig. \ref{Diff Rel Kin}c. Equation \ref{strain rate diffusion} predicts a power relation with an exponent between 1 and 2, observed experimentally \cite{Spiers1990,Gratier2009,Gratier2014,Berest2019}.
In agreement with those observations, data from Fig. \ref{Diff Rel Kin}c are well-fitted by a relation $\dot{y}=a\times P+b$. An even better fit has been found, considering $\dot{y}=a\times P^b$ (the exponent $b$ equals 1.4). Indeed the pressure applied $P$ is inside an Arrhenius exponential function in Equation \ref{Solid Activity Equation}. The linear dependency is only an approximation in a small pressure domain, even if the exponent (1.4) is similar to a linear one (1). A proportional fit has been tried, $\dot{y}=a\times P$, but it seems worse than the others because the linear dependence is only an approximation. The influence of the pressure applied on a bigger domain can hardly be verified experimentally as the pressure-solution is dominant in low pressure domain but phenomena like dislocation, plasticity or microcracking appear in large pressure one \cite{Urai2008}.

The influence of the indenter diameter $d$ ($\sim$ distance to the free fluid) on the strain rate $\dot{y}$ is explored in Fig. \ref{Diff Rel Kin}d. According to Equation \ref{strain rate diffusion} and experimental observations \cite{Spiers1990}, a relation $\dot{y}\propto 1/d^2$ is expected.
In agreement with those observations, data from Fig. \ref{Diff Rel Kin}d are well-fitted by a relation $\dot{y}=a/d^2$. 

The different data obtained in this Section are plotted together in Fig. \ref{Diff de_sim de_the}, representing $\dot{y}_{simulation}$ versus $\dot{y}_{theory}$. From Equation \ref{strain rate diffusion}, one can define $\dot{y}_{theory}=\alpha\;\frac{\kappa_cw\, c_{eq}\, P^{1.4}}{d^2}$, with $\alpha$ a coefficient describing a combination of the other parameters (well saturation, barrier energy, mobility, temperature, molar volume, examples among others). This coefficient $\alpha$ is estimated from the average of the different data. 
Then, the mean relative error with the target line $\dot{y}_{simulation}=\dot{y}_{theory}$ is computed. It appears the relative error of the simulated kinetic to the theoretical one is less than 4\% (with a maximum value of 12\%), a reflection of the simulations' consistency with the theoretical framework.

%%===========================%%

\subsection{Validation of the dissolution rate-limiting scenario}
\label{Validation Dissolution Section}

The second rate-limiting scenario is the dissolution one, defined as the dissolution process is the slowest \cite{Raj1982,Spiers1990b,vanNoort2008b,Gratier2009}. Numerically this scenario is verified by the criterion: $\overline{sat_{film}}\leq20\%$, where $\overline{sat_{film}}$ is the mean saturation in the fluid film. It is defined as $sat_{film}=\frac{c-c_{eq}}{a_s c_{eq}-c_{eq}}$, where $c$ is the solution concentration, $c_{eq}$ is the solution concentration at the equilibrium and $a_s$ is the solid activity defined Equation \ref{Solid Activity Equation}. 
The displacement rate $\dot{y}$ during dissolution rate-limiting steady state is defined at Equation \ref{strain rate dissolution} \cite{Raj1982,Spiers1990b,Gratier2009}.

\begin{equation}
    \dot{y}\propto \kappa_{diss}\, P^n
    \label{strain rate dissolution}
\end{equation}
where, $\kappa_{diss}$ is the dissolution kinetic, $P$ is the pressure applied and $n$ is an exponent applied to the pressure ($\approx 1-2$ \cite{Berest2019}). 

\begin{figure}[h]
    \centering
    a) \includegraphics[width=0.45\linewidth]{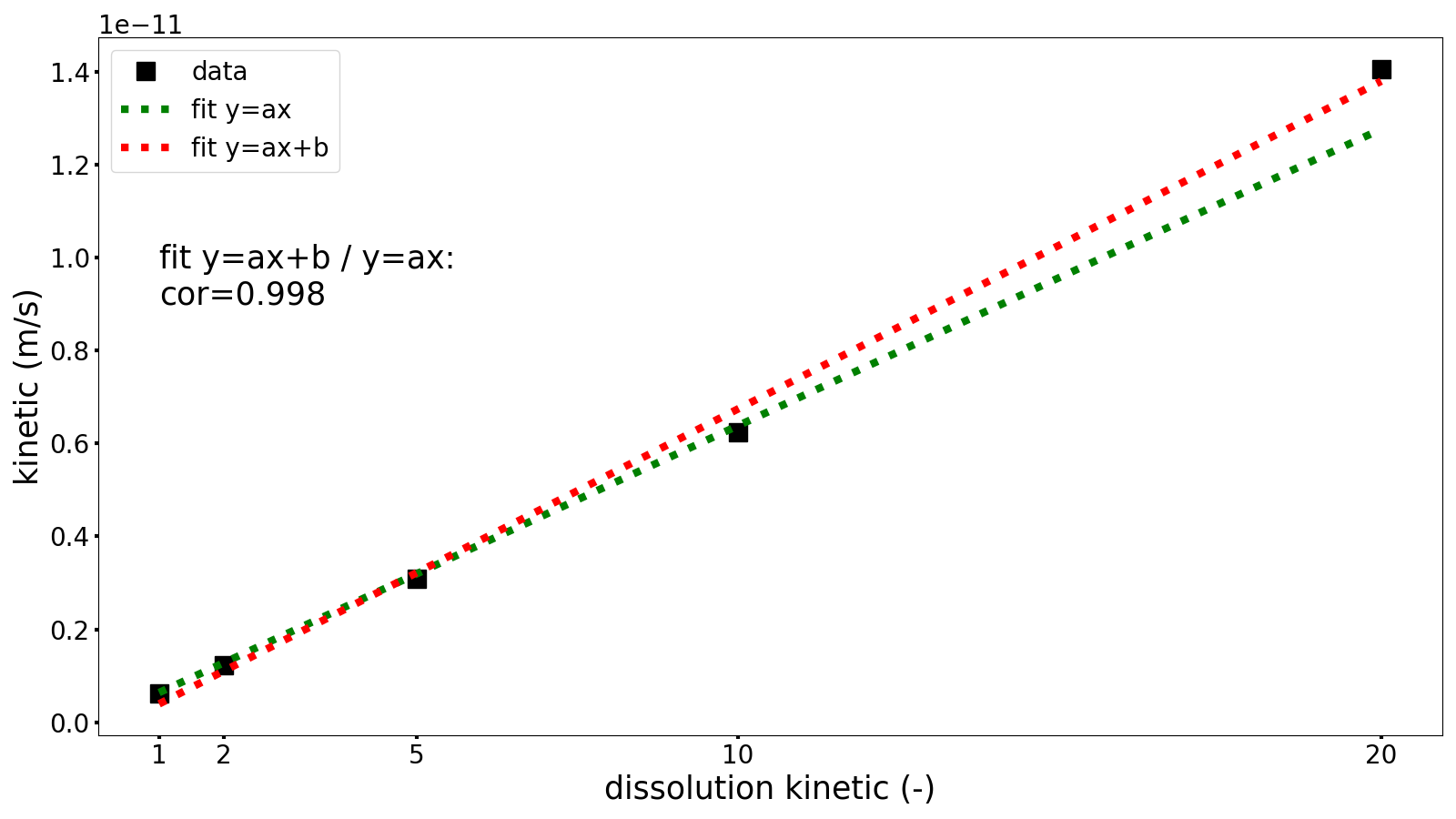}~
    b) \includegraphics[width=0.45\linewidth]{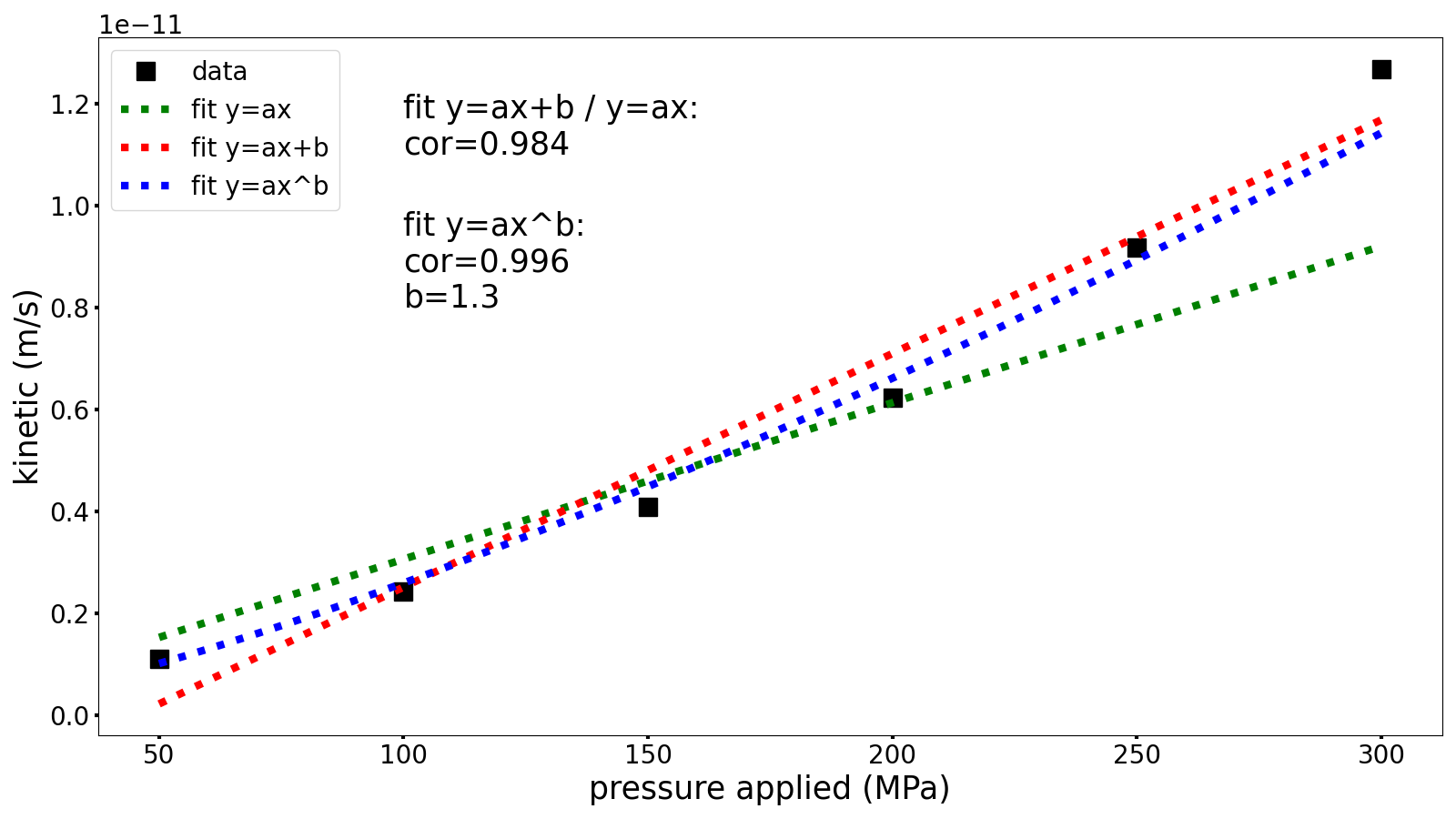}
    \caption{Comparison with the analytical model of the simulated influence on the kinetic of the indenter test of a) the dissolution kinetic and b) the pressure applied.}
    \label{Diss Rel Kin}
\end{figure}

\begin{figure}[h]
    \centering
    \includegraphics[width=0.6\linewidth]{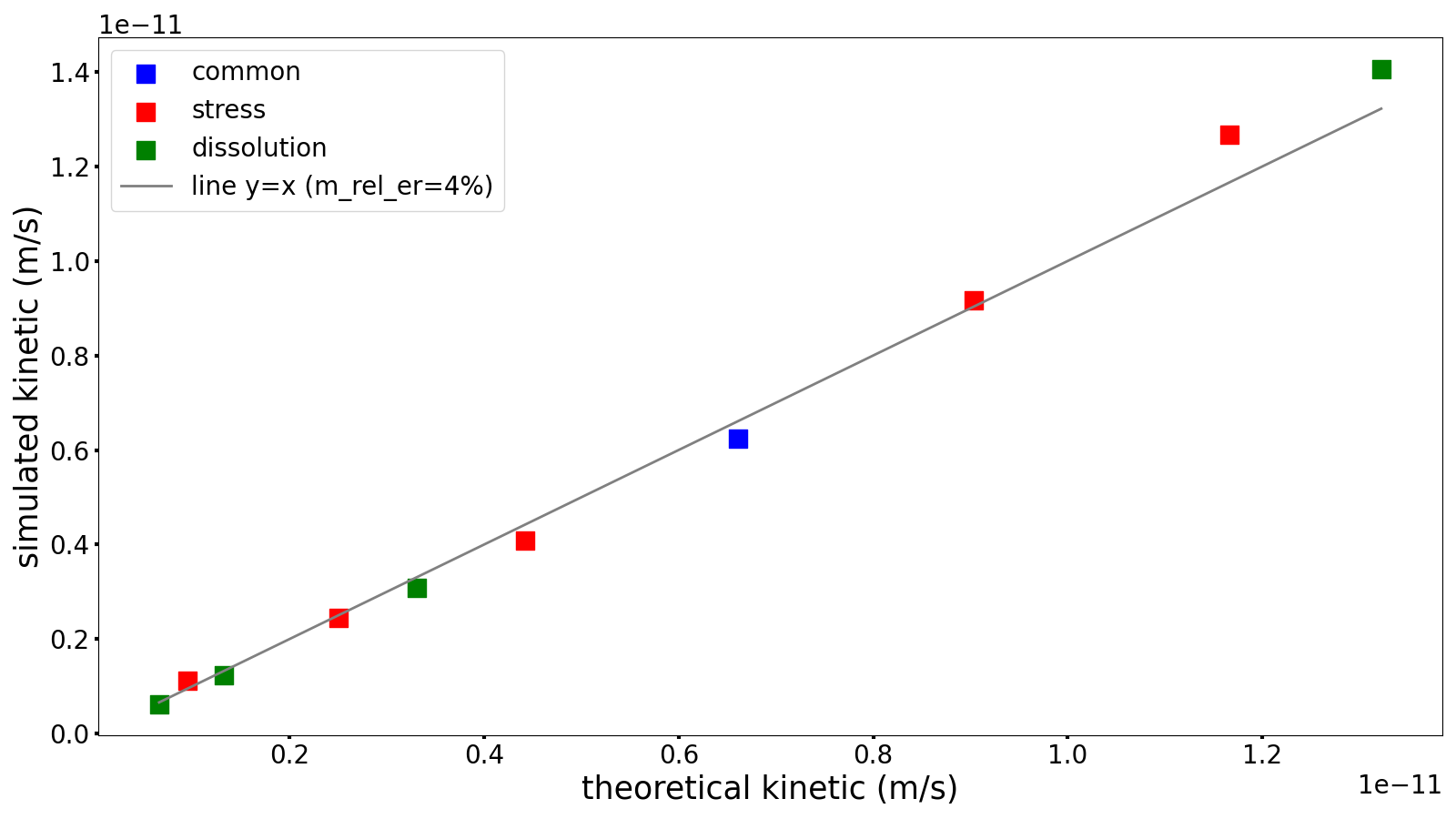}
    \caption{Comparison of the simulated and theoretical kinetics of the indenter test in the case of the dissolution rate-limiting.}
    \label{Diss de_sim de_the}
\end{figure}

Similar to the previous Section, a parametric investigation is conducted to verify the relation at Equation \ref{strain rate dissolution}. The parameters are shown in Table \ref{Parameter Indentation Dissolution}. To isolate the effect of each parameter, default values are considered (in bold), whereas a domain is explored in the specific investigation (default and domain definition from \cite{Gratier2009}). Results are given in Fig. \ref{Diss Rel Kin} and \ref{Diss de_sim de_the}.

\begin{table}[h]
    \centering
    \begin{tabular}{|l|r|}
        \hline
        dissolution kinetic $\kappa_{diss}$& 0.01-0.02-0.05-\textbf{0.1}-0.2\\
        pressure applied $P$& 50-100-150-\textbf{200}-250-300 MPa\\
        \hline
    \end{tabular}
    \caption{Parameters used to verify Equation \ref{strain rate dissolution}. Bold values represent the default ones, whereas the others represent the ones used only in the specific investigation.}
    \label{Parameter Indentation Dissolution}
\end{table}

The influence of the dissolution kinetic (through the coefficient $\kappa_{diss}$) on the strain rate $\dot{y}$ is explored in Fig. \ref{Diss Rel Kin}a. According to the Equation \ref{strain rate dissolution}, a proportional relation is expected. This factor has been observed experimentally \cite{DeMeer1997}.
In agreement with those observations, data from Fig. \ref{Diss Rel Kin}a are well-fitted by a relation $\dot{y}=a\times \kappa_{diss}$. 

The influence of the pressure applied $P$ on the strain rate $\dot{y}$ is explored in Fig. \ref{Diss Rel Kin}b. It is reasonable to posit a power relationship between the variables in question, see Equation \ref{strain rate dissolution} and experiments \cite{DeMeer1997,Niemeijer2002}.
In agreement with those observations, data from Fig. \ref{Diss Rel Kin}b are well-fitted by a relation $\dot{y}=a\times P+b$. Similar to the case of the diffusion rate-limiting, an even better fit has been found, considering $\dot{y}=a\times P^b$ (the exponent $b$ equals 1.3). As explained above, the pressure applied $P$ is inside an Arrhenius exponential function in Equation \ref{Solid Activity Equation}. The linear dependency is only an approximation in a small pressure domain, even if the exponent (1.3) is similar to a linear one (1). A proportional fit has been tried, $\dot{y}=a\times P$, but it seems worse than the others because the linear dependence is only an approximation. As explained above, the influence of the pressure for large values cannot be formed during experiments as other mechanisms are triggered \cite{Urai2008}.

Similar to the previous Section, the different data obtained in this Section are plotted together in Fig. \ref{Diss de_sim de_the}, representing $\dot{y}_{simulation}$ versus $\dot{y}_{theory}$. From Equation \ref{strain rate dissolution}, one can define $\dot{y}_{theory}~=~\alpha\;\kappa_{diss} P^{1.3}$, with $\alpha$ a coefficient describing a combination of the other parameters (well saturation, barrier energy, mobility, temperature, molar volume, examples among others). This coefficient $\alpha$ is estimated from the average of the different data. 
Then, the mean relative error with the target line $\dot{y}_{simulation}=\dot{y}_{theory}$ is computed. It appears also that the relative error of the simulated kinetic to the theoretical one is less than 4\% (with a maximum value of 11\%).

%%=======================================================%%

\section{Application to a granular configuration}

Once the formulation of the Phase-Field description used so far has been calibrated and validated, it can be coupled to a Discrete Element Model to capture the heterogeneous dissolution/precipitation and the granular reorganization \cite{SacMorane2024}. In fact, the phase-field formulation itself is not able to compute the granular mechanical steady-state and interpolate the stress at the contacts.
In the following Sections, the formulation of the discrete element model and its coupling with the phase-field description are abbreviated. In fact, some modifications have been made compared to \cite{SacMorane2024}. In particular, the contact law between particles has been modified, from an overlap-based to a volume-based one. This point is primordial to face undesirable properties as the fact that a contact force can be computed equal for two contacts with the same overlap but different overlapping volumes \cite{vanderHaven2023,Feng2023}. 

Here, a new numerical configuration is considered: a constant force is applied to a grain subject to pressure-solution at the contact with a plate not subject to pressure-solution. Compared to the previous indenter test, the material subject to pressure-solution moves and the pressure at the contact is initially unknown as the contact surface evolves. 
Particular attention is paid to the influence of precipitation on the phenomenon kinetic. In fact, precipitation induces an increase of the contact surface. It reduces the stress transmitted at the contact level. And, it produces a slowdown of the creep settlement. In the currently applied models at the microscale \cite{Spiers1990b,Lu2021}, this effect of the precipitation is not considered.

%%===========================%%

\subsection{DEM formulation}
\label{DEM Section}

The Discrete Element Method has been developed to account for the individual grains and their interactions \cite{Cundall1979,OSullivan2011}. The momentum balances are solved for each grain, following different integration methods \cite{Samiei2013}. To capture the heterogeneous shape, a polyhedral description is used \cite{Cundall1988,Nezami2004,Latham2004,Kawamoto2016}. This simulation part is solved thanks to the open-source software Yade \cite{YADE}.

The only difference with \cite{SacMorane2024} is the fact that the contact law between particles has been modified, from a nonlinear overlap-based one to a linear volume-based one \cite{vanderHaven2023,Feng2023}.

%%===========================%%

\subsection{Couplings between Phase-Field and DEM}
\label{PF-DEM Couplings Section}

The data exchange and global scheme of the PF-DEM couplings are explained in \cite{SacMorane2024}. 
In summary, the phase-field simulation entails the discretization of the sample geometry into a mesh and the representations of grains by phase variable maps.
A grain detection algorithm is employed to ascertain the novel grain shape and solute configuration, based on the phase-field outputs. 
In contrast, the discrete element modelization employs a representation of grains as polygonal particles. 
The objective of this model is to compute the new positions of the grains and the pressure applied at each contact point.
It is important to note that although PF and DEM operate on disparate time scales, they are nevertheless related. The mechanical equilibrium (DEM) is assumed to occur instantaneously, whereas grain dissolution or precipitation (PF) occurs over longer time periods, typically spanning several hours to days.

%\begin{figure}[ht]
%\centering
%\includegraphics[width=0.4\linewidth]{pf_to_dem.png}
%\caption{Once a mesh element is eligible (at least one node $\eta<0.5$ and one node $\eta>0.5$), a vertex is computed. First interpolations are done on sides crossed by the isoline $\eta=0.5$. Then the vertex is computed, considering the average coordinates of them.}
%\label{PFtoDEM Figure}
%\end{figure}

%A polygonal particle is obtained based on the vertex positions. To reduce computational cost, only a certain number of vertices are used.
%As explained in Fig. \ref{Grains List Vertices}a, all nodes are considered initially. Then, the grain is divided into equal angular parts and a mean vertex is computed for each division, see Fig. \ref{Grains List Vertices}b. This operation helps to save computational cost by reducing the number of vertices. As explained in \cite{SacMorane2024}, the quality of the result depends on this number of vertices. $20-30$ nodes seem a good compromise.
%Finally, the geometric properties of the grain, such as its center, surface area, and inertia, can be computed from the list of vertices.

%\begin{figure}[ht]
%    \centering
%    a) \includegraphics[width=0.2\linewidth]{From_PF_to_DEM_2a.png}~
%    b) \includegraphics[width=0.2\linewidth]{From_PF_to_DEM_2b.png}
%    \caption{Even if a) the grain is described as a list of vertices (in black), b) only a part of this list is considered by computing mean vertices (in red) to save computational cost.}
%    \label{Grains List Vertices}
%\end{figure}

Once the grain boundaries are identified, a series of DEM iterations are conducted until the system reaches an equilibrium. To dissipate the mechanical energy, a certain degree of damping is introduced. The convergence criterion for detecting the equilibrium may vary depending on the specific problem under consideration. In this study, equilibrium is identified when the applied force falls within the range of 0.98 to 1.02 times the target force over a period of 100 iterations.
Furthermore, a maximum number of iterations is defined to prevent an infinite loop. In the event that equilibrium is reached prior to the maximum number of iterations, new phase maps are constructed in order to update the geometry of the phase field problem based on the results of the DEM step.

The new phase map is obtained through interpolation from the previous one. As illustrated in Figure 13 of \cite{SacMorane2024}, at each iteration, the subsequent mesh is computed based on the preceding mesh through the application of a rigid body motion. As the two meshes frequently exhibit imperfect alignment, bilinear (2D) or trilinear (3D) interpolation is employed to estimate the values of the subsequent mesh.

\begin{figure}[ht]
\centering
\includegraphics[width = 0.5\linewidth]{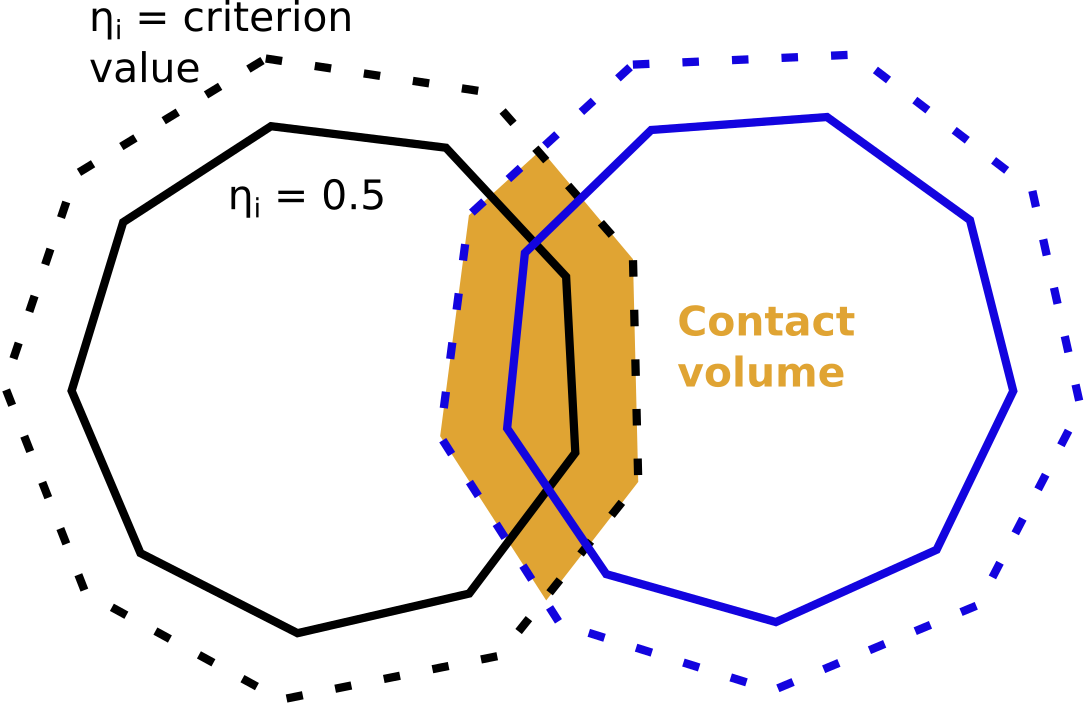}
\caption{Definition of the contact volume between two grains: a criterion value lower than $\eta_i=0.5$ can be used to enlarge the part of the particle under the solid activity $a_s$ increase.}
\label{Contact Volume}
\end{figure}

Prior to initiating a novel phase-field simulation, it is essential to compute the tilting term $E_d$ through the solid activity $a_s$, see Equation \ref{Tilting Term}. In the previous Section, $a_s$ is readily computed given that the pressure applied at the contact is known and constant. However, this is not the case in the present Section. Indeed, the pressure applied at the contact evolves with the shape evolution of the grains and the granular reorganization (in the case of multiple grains). The pressure is then interpolated from the DEM simulation, $P=F_n/S_c$.
The applied pressure depends on the normal force $F_n$ transmitted within the DEM contact and the contact surface $S_c$. The contact surface is computed based on the contact volume $V_c$. As illustrated in Fig. \ref{Contact Volume}, the contact volume is determined by iteration on the mesh nodes. The criteria for detecting contact is $\eta_i$ and $\eta_j\geq\eta_{criterion}$, where $\eta_{i,j}$ are the phases linked to the grains in contact and $\eta_{criterion}$ is a criterion value. The initial approach is to assume that $\eta_{criterion}=0.5$ as it represents the grain boundaries. Nevertheless, some approaches could consider $\eta_{criterion}<0.5$. This is to account for the possibility that a supplementary part of the grain participates in the contact. In the following Section, we assume that $\eta_{criterion}=0.1$.

Once the contact volume $V_c$ has been determined, the contact surface $S_c$ is interpolated. To this end, the maximum orthogonal line (2D) or plane (3D) to the normal contact vector $\overrightarrow{n}$ employed in the DEM part is sought. Once this parameter has been determined, the solid activity $a_s$ is calculated, see Equation \ref{Solid Activity Equation}. It should be noted that $a_s=1$ out of the contact volume $V_c$. Finally, the tilting term $E_d$ is computed, see Equation \ref{Tilting Term}.

%%===========================%%

\subsection{Numerical model}

\begin{figure}[h]
    \centering
    \includegraphics[width=0.5\linewidth]{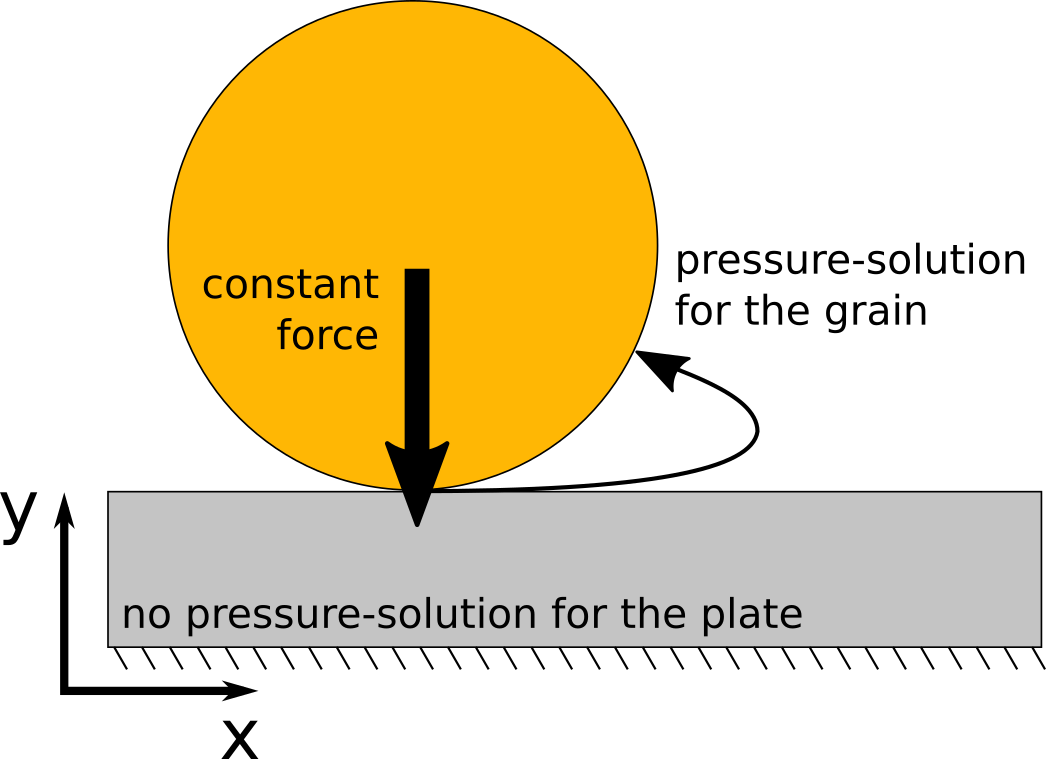}
    \caption{A constant force is applied to a grain subject to pressure-solution over a plate not subject to pressure-solution.}
    \label{Grain Plate Pressure Solution Figure}
\end{figure}

Fig. \ref{Grain Plate Pressure Solution Figure} depicts the configuration investigated in this Section. A constant force is applied to a grain that is subject to pressure-solution over a plate that is not subject to pressure-solution. The material undergoes dissolution at the contact, subsequent diffusion within the fluid film, and finally precipitation on the surface less stressed. Table \ref{Grain Plate Pressure Solution Table} presents the relevant parameters utilized in this study. The majority of these parameters are consistent with those employed in the calibration and validation campaigns detailed in the preceding Sections. In this instance, particular attention is directed towards the impact of precipitation kinetics coefficient $k_{prec}$ on creep behaviour. Indeed, this impact on creep behavior is not considered in the current models. It should be noted that, in contrast to previous simulations, the solute concentration is not evacuated in this instance, as the domain is closed.

\begin{table}[h]
    \centering
    \begin{tabular}{|l|l|}
        \hline
        Initial radius & $100\; \mu$m\\
        Thickness of the simulation & $100\; \mu$m\\
        Volume overlap / Initial grain volume & $1.6\;$\% \\
        \hline
        Mobility $L_j$ & $3\;$s$^{-1}$\\
        Mesh size $\Delta x$ & Initial radius / 40 \\
        Dissolution kinetic $k_{diss}$ & $0.005/\Delta x$\\
         & $0.5-1-2-5-$\\
        Precipitation kinetic $k_{prec}$ & $8-10-13-16\; k_{diss}$\\
        \hline 
        Equilibrium constant $c_{eq}$&\\
        Molar volume $V_{s}$&\\
        Gas constant $R$& same as Table \ref{Parameter Known}\\
        Temperature $T$&\\
        Diffusivity $\kappa_{c}$ $\times$ film width $w$&\\
        \hline
    \end{tabular}
    \caption{Parameters used for the grain-plate configuration.}
    \label{Grain Plate Pressure Solution Table}
\end{table}

%%===========================%%

\subsection{Results}

\begin{figure}[h]
    \centering
    \includegraphics[width=0.6\linewidth]{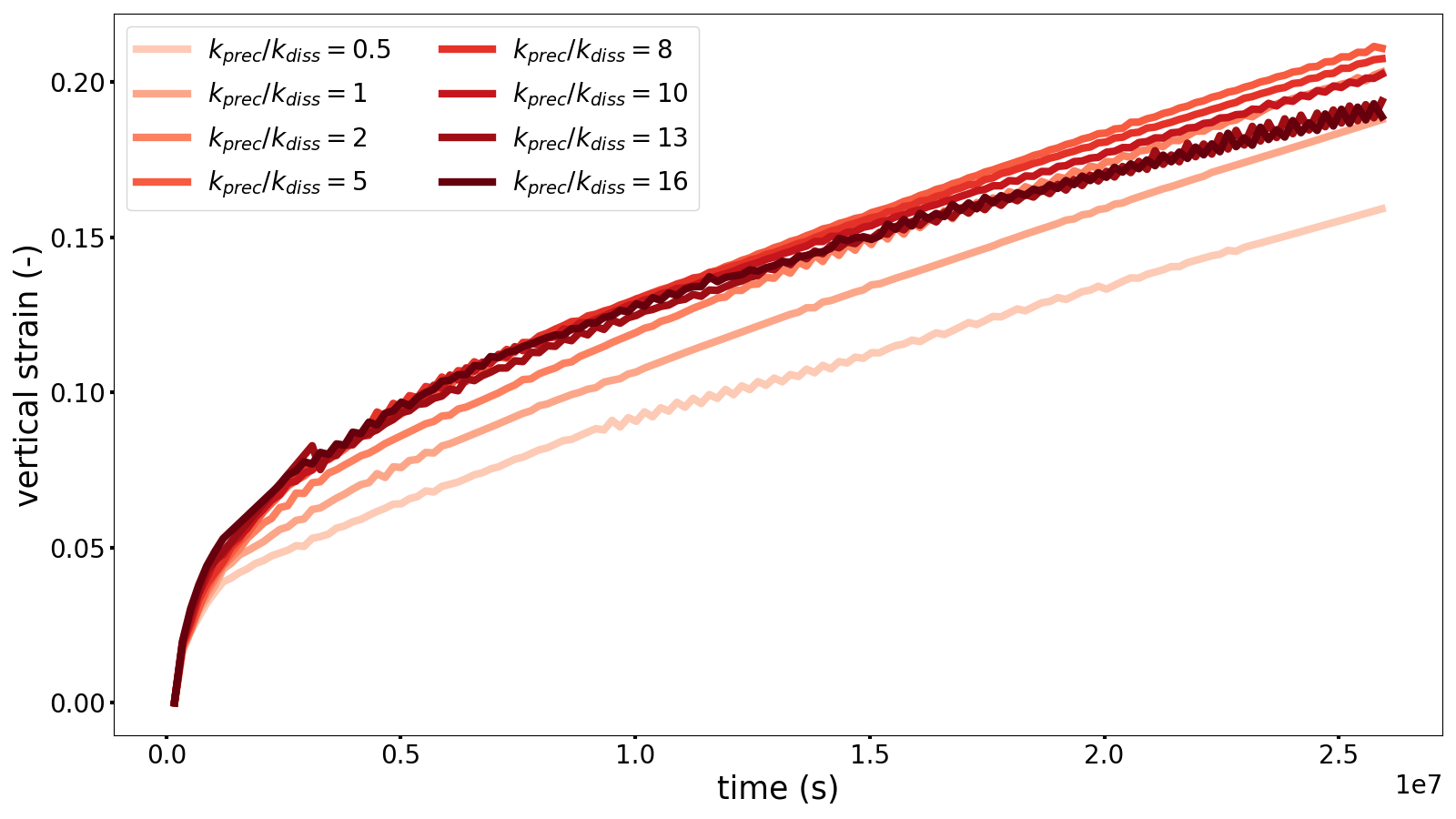}
    \caption{Time evolution of the vertical strain for different precipitation kinetics.}
    \label{Strain Time Precipitation}
\end{figure}

Fig. \ref{Strain Time Precipitation} presents the vertical strain's time evolution for distinct precipitation kinetic and a movie example is provided as a supplemental material. The definition of the vertical strain is $\epsilon=\Delta y/2\,R$, where $\Delta y$ is the cumulative vertical displacement of the grain and $R$ is the initial radius.

As detailed in \cite{Spiers1990,Spiers1990b}, the settlement kinetic can be estimated, see Equation \ref{Spiers Equation}, under the assumption of a close-packed (face-centered-cubic lattice) array of spherical grains with identical diameters. Equation \ref{Spiers Equation Solution} provides the solution to this partial differential equation, which is a $n+1$ root law. The exponent $n$ determines the extent of the slowdown in creep behavior; a larger value of $n$ corresponds to a stronger slowdown.

\begin{equation}
    \dot{\epsilon} = k \; 1/\epsilon^n
    \label{Spiers Equation}
\end{equation}
where, $k$ contains the different factors such as the pressure applied, the grain diameter, and the diffusivity of the solute, among others.

\begin{equation}
    \epsilon(t) = \left(n\,k\,t + k\,t \right)^{1/(n+1)}
    \label{Spiers Equation Solution}
\end{equation}

\begin{figure}[h]
    \centering
    \includegraphics[width=0.6\linewidth]{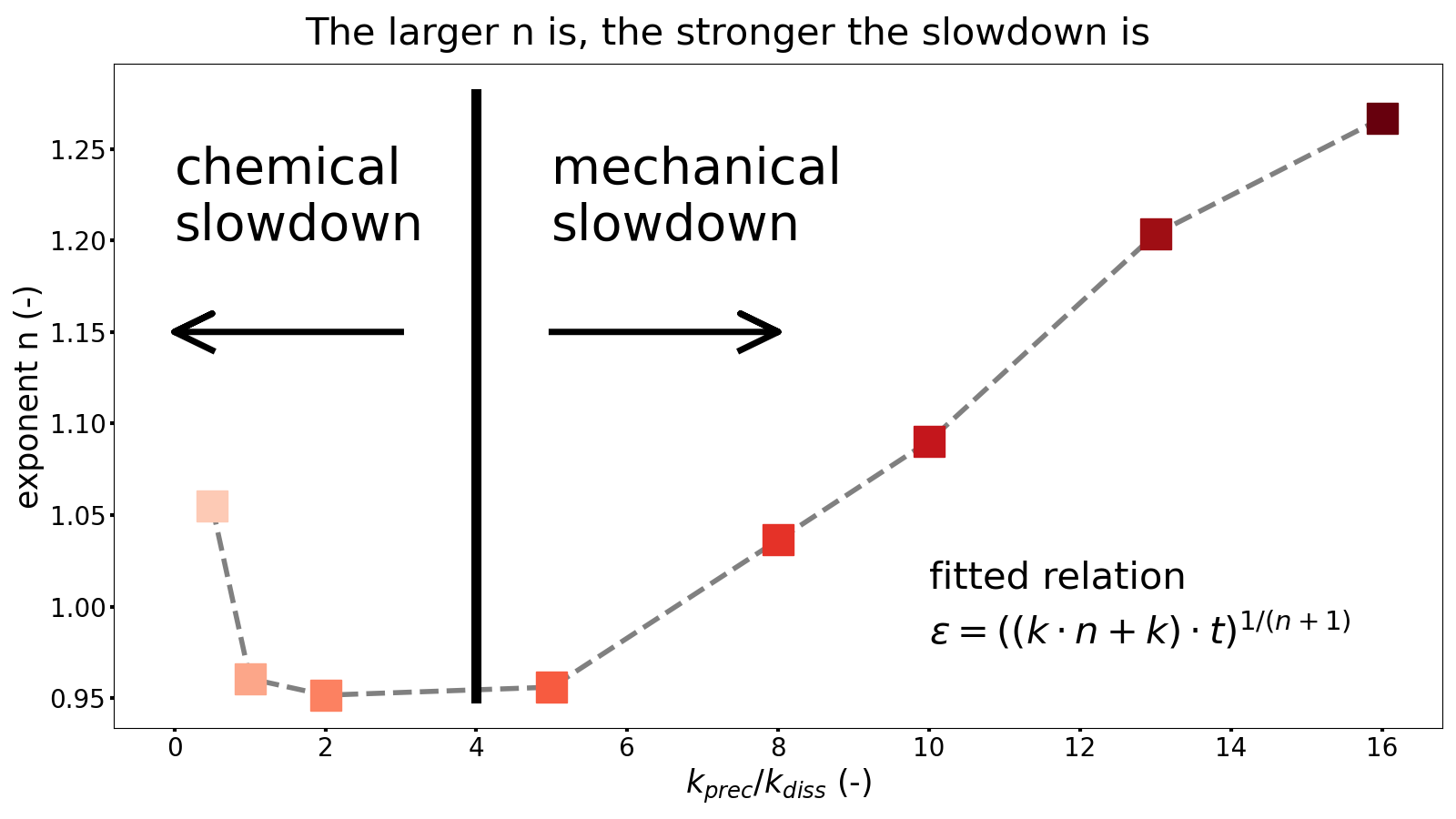}
    \caption{Influence of the precipitation kinetic on the slowdown (through the exponent $n$) of the evolution of the vertical strain $\epsilon(t)$.}
    \label{Exponent State Precipitation}
\end{figure}

This relationship is employed to compute a fit of the results obtained. It appears the fit provides a correlation coefficient of approximately $0.997$ for the various curves. The factor $n+1$ appears to be approximately $1.95$ and $2.25$, which is consistent with the experimental observations in \cite{Spiers1990}.
Subsequently, the impact of the precipitation kinetic on the exponent $n$ can be discerned in Fig. \ref{Exponent State Precipitation}. 

It seems that at low precipitation rates $\left(k_{prec}/k_{diss} \leq 4\right)$, the slowdown declines in proportion to the precipitation kinetic. Conversely, the slowdown increases with the precipitation kinetic at fast precipitation $\left(k_{prec}/k_{diss} \geq 4\right)$. The slowdown of the pressure-solution creep is caused by (i) a chemical mechanism or (ii) a mechanical one \cite{Lu2021}. Indeed, if the solute concentration builds up in the pore space, the mass transfer by diffusion is slowed down, case (i). Similarly, an increase in the contact surface results in a reduction in pressure (the motor of pressure-solution), and the mass flux by dissolution is slowed down, case (ii) \cite{Lu2021}.

\begin{figure}[h]
    \centering
    \includegraphics[width=0.6\linewidth]{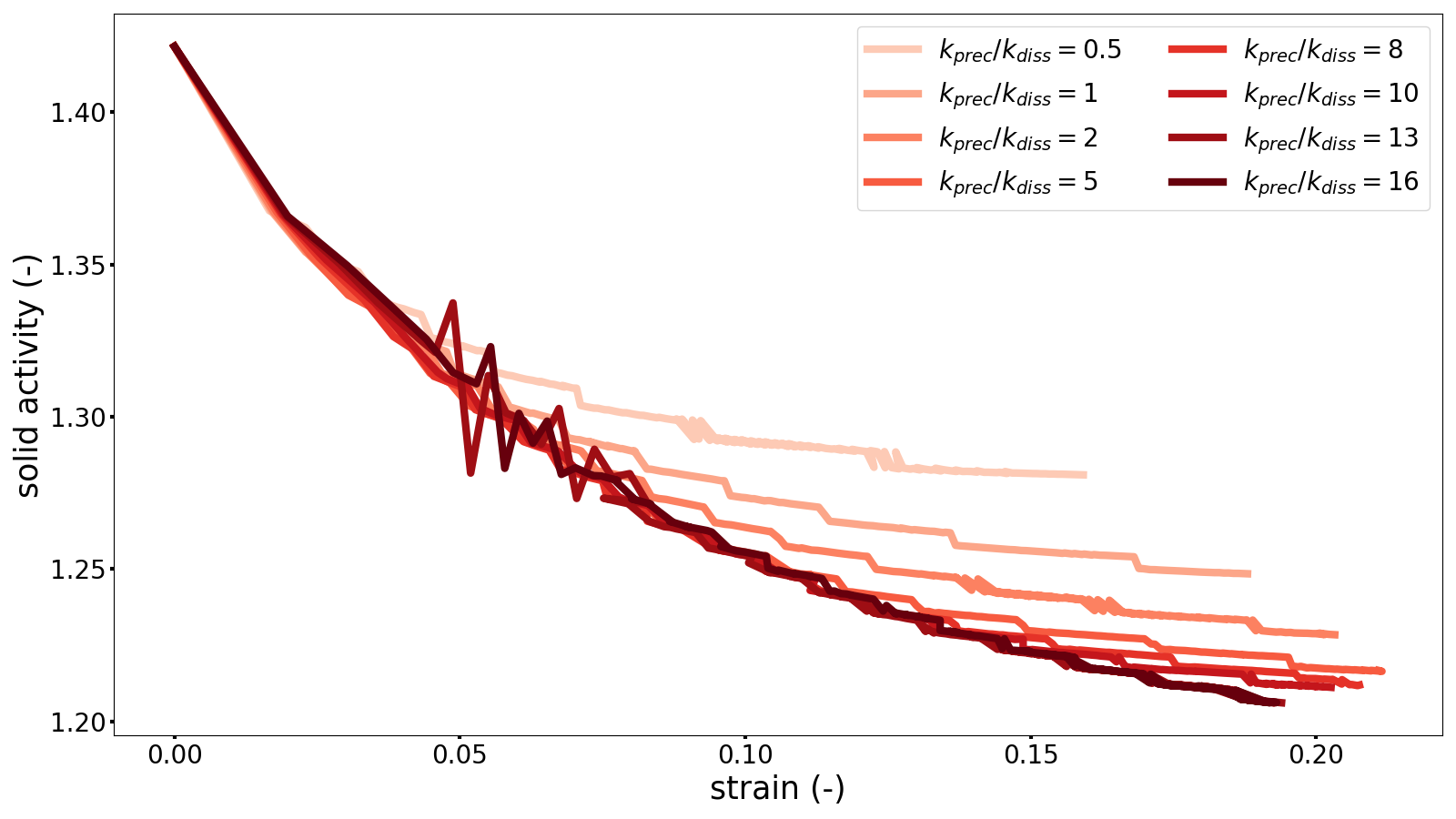}
    \caption{Influence of the precipitation kinetic on the solid activity ($\sim$ pressure at the contact) during the creep behavior. At a given strain, the solid activity decreases with precipitation kinetics.}
    \label{Solid Activity Strain Precipitation}
\end{figure}

Fig. \ref{Solid Activity Strain Precipitation} illustrates the evolution of the solid activity ($\sim$ pressure at the contact) in relation to the vertical strain for varying precipitation kinetics. The impact of the precipitation kinetic is readily discernible. As expected, the solid activity is observed to be smaller at a given strain for faster precipitation kinetics. It demonstrates the slowdown of the pressure-solution due to a mechanical mechanism.

\begin{figure}[h]
    \centering
    \includegraphics[width=0.6\linewidth]{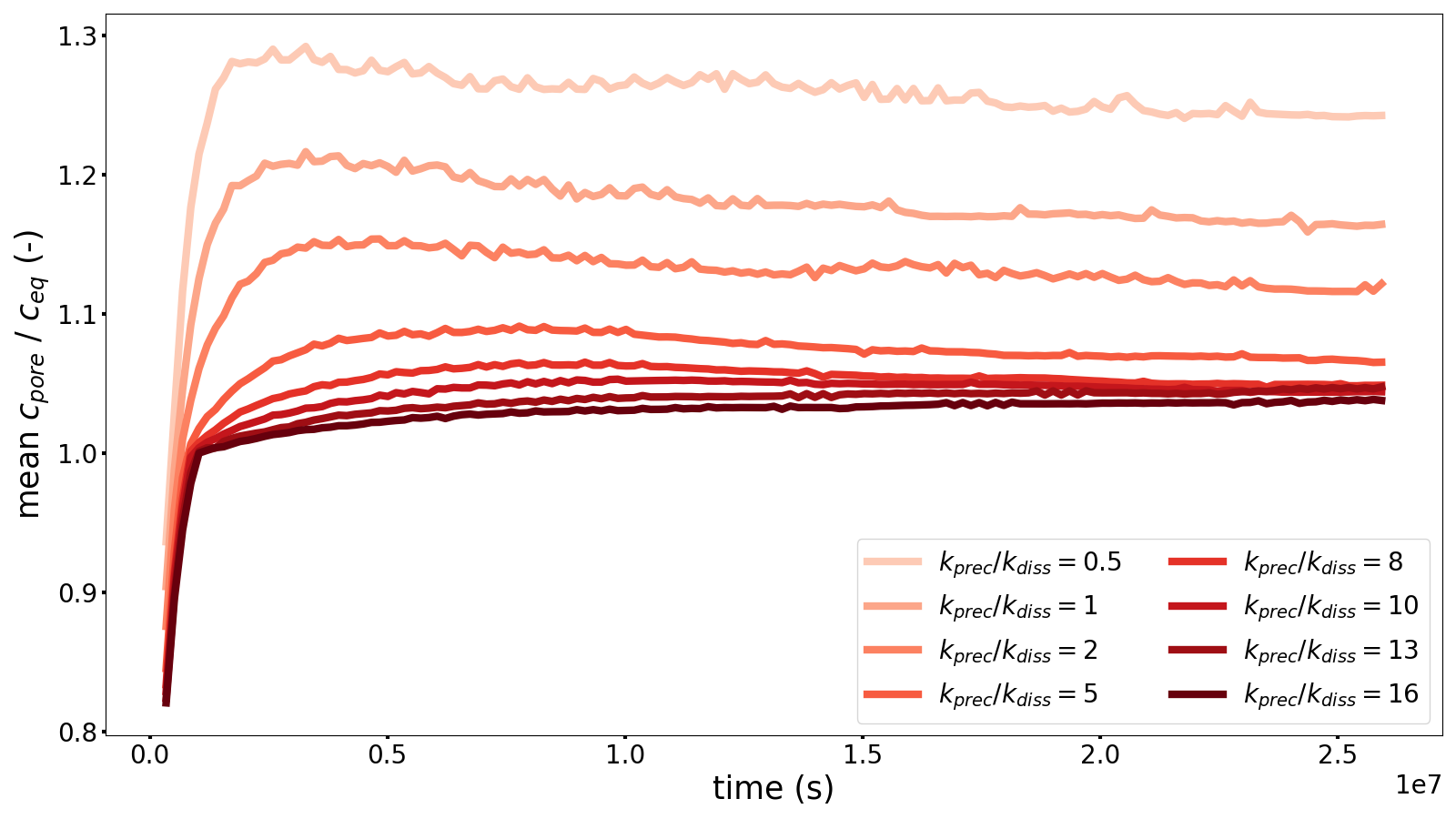}
    \caption{Influence of the precipitation kinetic on the mean value of the solute concentration in the pore space during the creep behavior. As precipitation kinetic increases, the accumulation of the solute concentration in the pore space decreases.}
    \label{C Pore Time Precipitation}
\end{figure}

Nevertheless, Fig. \ref{Exponent State Precipitation} underscores a slowdown in the creep behavior, even when the precipitation kinetic decreases. This is particularly evident in the case of slow precipitation $\left(k_{prec}/k_{diss} \leq 4\right)$.
This is contrary to the indications provided by Fig. \ref{Solid Activity Strain Precipitation}.
Fig. \ref{C Pore Time Precipitation} illustrates the evolution of the mean value of the solute concentration (in comparison to the equilibrium value) within the pore space for varying precipitation kinetics coefficient. The solute concentration appears to build up in the pore space for slow precipitation. The slowdown of the pressure-solution is attributed to a chemical mechanism.

\begin{figure}[h]
    \centering
    \includegraphics[width=0.7\linewidth]{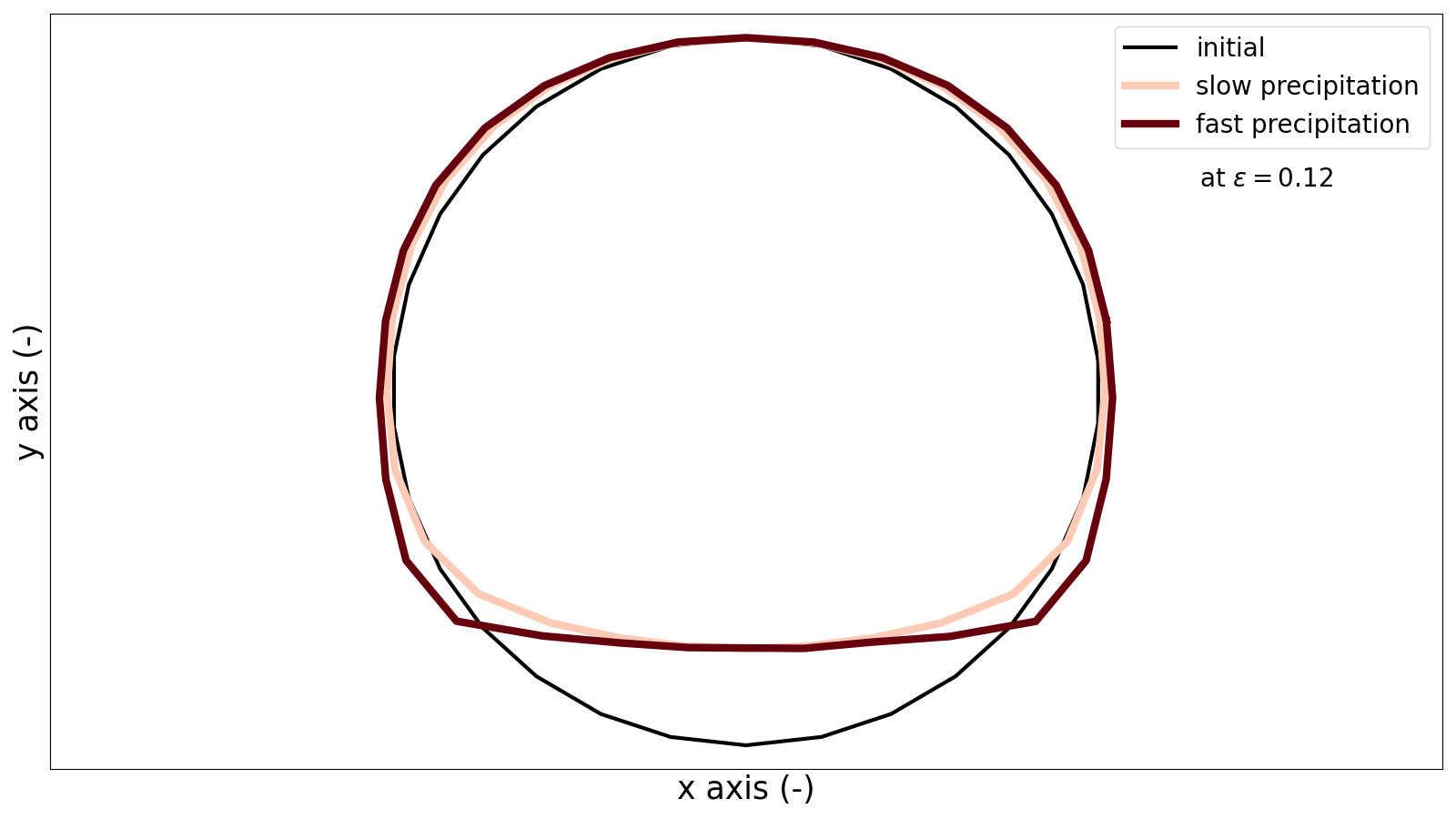}
    \caption{Comparison of the initial grain shape with the current one at a given strain for slow and fast precipitation kinetics. The contact surface increases with the precipitation kinetic.}
    \label{Grain Shape Precipitation}
\end{figure}

Upon returning to Fig. \ref{Exponent State Precipitation}, two distinct domains become evident. In the case of slow precipitation $\left(k_{prec}/k_{diss} \leq 4\right)$, the creep slowdown (due to chemistry) decreases in line with the precipitation kinetic. In contrast, for fast precipitation $\left(k_{prec}/k_{diss} \geq 4\right)$, the creep slowdown (due to mechanic) increases with the precipitation kinetic. As observed by \cite{Lu2021} through a 1D model, these two limiting scenarios can be seen to occur when the solute outflux from the contact zone is either prevented or allowed \cite{Lu2021}. 
Preventing the solute outflux results in the accumulation of the solute concentration in the pore space. It is a chemical slowdown. Conversely, allowing the solute outflux results in the increase of the surface contact, as the dissolution occurs. However, the model of \cite{Lu2021} does not account for the impact of the precipitation in the increase of the contact surface or the consumption of the solute concentration in the pore space.
Precipitation was included in the model proposed by \textit{Yasuhara et al.} in 2003 \cite{Yasuhara2003}. However, their simplifying assumptions on the geometry and processes employed hinder the effect of precipitation on the strain rate. Nevertheless, the model allows for the observation of the impact of the precipitation on the porosity evolution.
Similarly, \textit{Xu and Arson} considered the precipitation as a healing process \cite{Xu2022}. The consequence of this phenomenon is an increase in the stiffness of the sample at the continuum scale. However, this model does not account for granular reorganization.
Additionally, it has been suggested that the precipitation kinetic can control the global creep behavior if the process is the slowest \cite{Mullis1991}. Nevertheless, this assessment is only based on the rate-limiting theory and does not consider the evolution of the grain shapes and configuration. 
The PFDEM framework presented here enable to solve for the muti-physical processes in the whole microstructure geometry and is also capable of capturing the evolution of the grain shape, see Fig. \ref{Grain Shape Precipitation}. The shape of the grain seems to depend on the precipitation kinetic. In particular, the contact surface increases in proportion to the precipitation kinetic. The complex geometries of the grains exert a significant influence on the macroscopic mechanical behavior of granular materials \cite{Binaree2019,Guevel2022} and need to be considered.

The consideration of the precipitation for pressure-solution kinetic, often neglected in the current models, is a crucial factor. It determines the slowdown mechanism (chemical or mechanical) and the strength of the slowdown, which varies depending on the precipitation kinetic. 

%%=======================================================%%

\section{Conclusion}

The Phase-Field Discrete Element Model framework has been employed to examine the influence of precipitation on the slowdown of creep behavior induced by pressure-solution. It is noteworthy that this particular aspect is not currently included in the models that are employed to describe the strain rate due to pressure-solution or their simplifying assumptions on the microstructure hinder its influence. To this end, the parameters were first calibrated using data from previous indentation experiments and the formulation has been validated by reproducing the well-known relations that are typically employed in rate-limiting scenarios. 
 Then, the model is applied to a grain-to-grain configuration and adding the precipitation process. At low precipitation rates, the slowdown rate declines in proportion to the kinetic factor. The underlying mechanism is linked to the accumulation of solute concentration within the pore space. Conversely, the slowdown increases with the kinetic at fast precipitation. In this case, the mechanism is due to the larger surface area at the contact. It decreases the stress, which is the driving force behind the phenomenon.

This article illustrates the necessity of integrating all relevant processes in the case of multi-physics phenomena, such as pressure-solution, but also solve them numerically at the microstructure scale. Indeed, this chemo-mechanical phenomenon is subdivided into three distinct processes: dissolution, diffusion, and precipitation and the global kinetic of this creep behavior is controlled by the slowest rate-limiting process. However, the rate-limiting process may undergo changes over time as the microstructure configuration is modified.
For instance, the rate-limiting process for pressure-solution is frequently dissolution during the initial stage. Indeed, the diffusion length is relatively short, and the solute dissolved in the fluid film evacuates rapidly into the pore space. However, the diffusion length increases in proportion to the material dissolution and precipitation around the contact. At a certain point, the diffusion may become the rate-limiting process. This transition can only be captured if models integrate a sufficient number of phenomena. Finally, this lays the foundation for a deeper understanding of pressure-solution at the microscale, enabling to investigate in the future complex microstructures, but also its role in diagenesis and earthquake nucleation for various minerals and environmental conditions. 

%%=======================================================%%

\begin{appendix}

%%=======================================================%%

\section{Preliminary studies of the modelization of the indenter test}

This Section illustrates the preliminary investigations realized before the simulation presented in the Section Model calibration through indentation experiments. The influence of several parameters is explored to ensure the quality of the results.

%%===========================%%

\subsection{Update frequency and well size dependency}
\label{Preliminary Frequency Well Size Dependency}

The update frequency is one of the most important parameters for the answer quality. Indeed, the solute in the well needs to be evacuated frequently (if not, the solute concentration aims to reach the saturation value and the diffusion kinetic slows down) and the diffusion map needs to be updated to follow the front evolution.

\begin{figure}[ht]
\centering
\includegraphics[width = 0.7\linewidth]{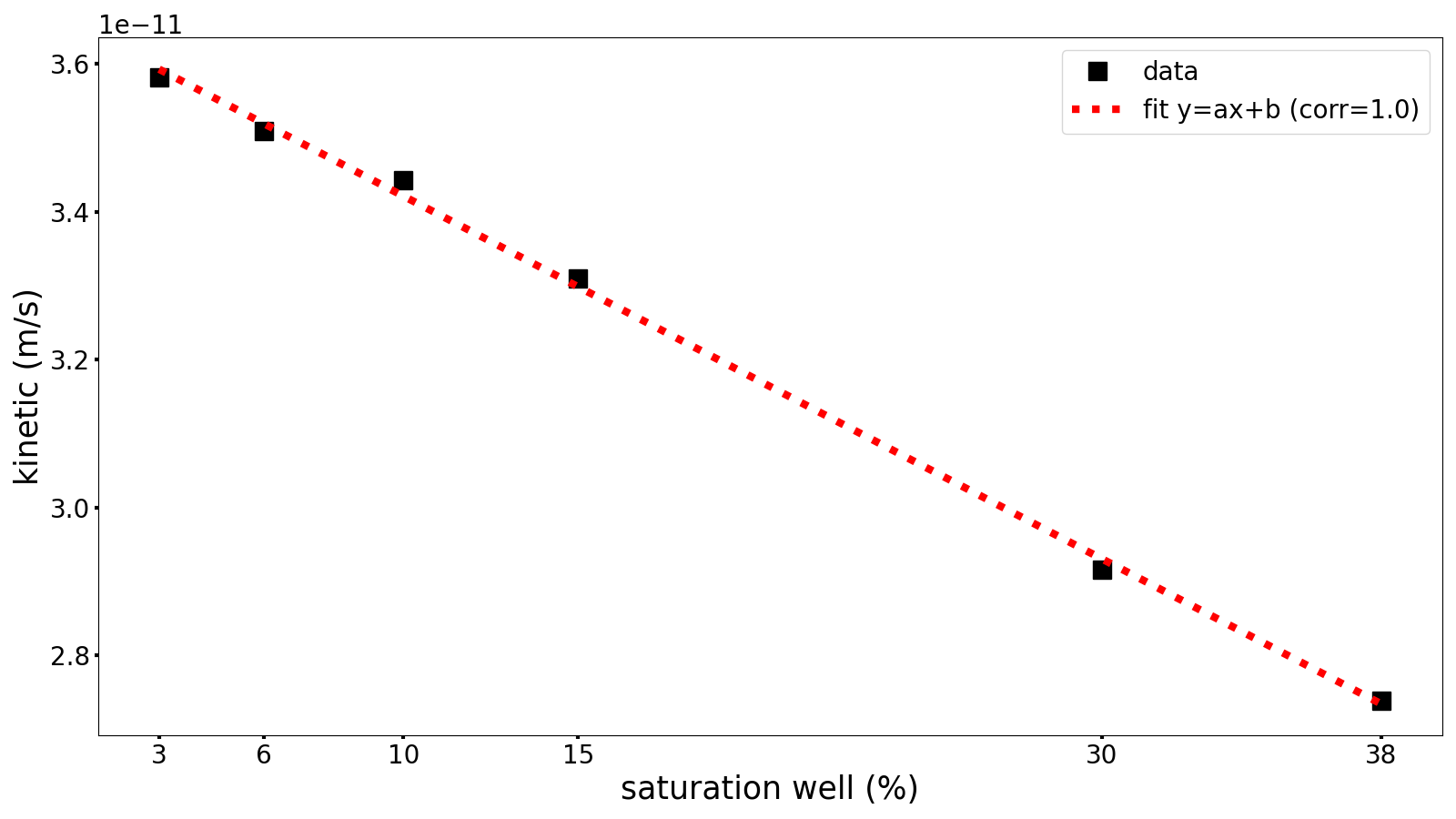}
\caption{Influence of the saturation well on the kinetic of the indenter test. An effort is needed to compare simulations at the same $sat_{well}$.}
\label{Diff Sat Kin}
\end{figure}

The effect of the saturation of the well $sat_{well}$ is explored in Fig. \ref{Diff Sat Kin}. A slowdown of the kinetic is well observed. Hence, it is important in Section Model calibration through indentation experiments to compare simulations with the same saturation well. It can be controlled by changing the map update frequency or adapting the well's size. 

%%===========================%%

\subsection{Mesh dependency}
\label{Preliminary Mesh Dependency}

As with all the finite element resolutions, the mesh should be small enough. A special focus is needed in the phase-field description presented in this article on the kinetic of the interface dissolution/precipitation. Indeed, it appears the interface dissolution/precipitation kinetic is related to \cite{Chan1977,Jou1997} :
\begin{itemize}
    \item the tilting amplitude $e_d$, see Fig. \ref{Destabilized Double Well} and Equation \ref{Tilting Term}
    \item the gradient coefficient $\kappa$ used in the Allen-Cahn formulation, Equation~\ref{AC formulation}. This coefficient is related to the mesh size, see Equation \ref{PF Parameters link}
\end{itemize}

Two methods seem possible to conserve the kinetic constant with various meshes: adapt the number of nodes defining the interface width to have a constant gradient coefficient $\kappa$ or adapt the tilting amplitude $e_d$. 
As the philosophy of the phase-field description is to keep a small number of nodes used for interface \cite{Takaki2014}, to approximate a sharp interface, the latter proposition is favored.

Considering the relation $e_d^j=e_d^i\times(m^i/m^j)$, the kinetic is conserved with different meshes, with $e_d^k$ and $m^k$ are the tilting term and the mesh size for the configuration $k$, The gradient coefficient $\kappa$ is computed with a constant number of mesh nodes.

\end{appendix}

%%=======================================================%%

\section{Data Availability Statement}

Some or all data, models, or code generated or used during the study are available in a repository online in accordance with funder data retention policies (\url{https://github.com/AlexSacMorane/PF_PS_Indentation} and  \url{https://github.com/AlexSacMorane/MY_PFDEM_1G_2D}). Some or all data, models, or code that support the findings of this study are available from the corresponding author upon reasonable request.

%%=======================================================%%

\section{Acknowledgements}

This research has been partially funded by the Fonds Spécial de Recherche (FSR), Wallonia-Bruxelles Federation, Belgium. The work has also received funding from the National Science Foundation (NSF), USA project CMMI-2042325. 

We would like to thank J.P. Gratier for his time and the discussion we had about the pressure-solution phenomenon.

%%=======================================================%%

\end{document}